\documentclass[longauth]{aa}  

\usepackage{graphicx}
\usepackage{txfonts}
\usepackage[]{hyperref}

\defcitealias{2020Sci...369.1347}{M20}

\hypersetup{
    colorlinks=true,
    citecolor=blue,
    linkcolor=blue,
    filecolor=magenta,      
    urlcolor=cyan,
}
\titlerunning{Galaxy-galaxy strong lensing in simulated galaxy clusters}
\authorrunning{Meneghetti et al.}
\begin{document}

   \title{The probability of galaxy-galaxy strong lensing events in hydrodynamical simulations of galaxy clusters}

   %\subtitle{I. Theoretical estimates based on different suites of hydrodynamical simulations}

    \author{Massimo Meneghetti\inst{\ref{oas},\ref{infnbo}}\thanks{\email{massimo.meneghetti@inaf.it}}, 
    Antonio Ragagnin\inst{\ref{unibo},\ref{oats},\ref{ifpu}},
    Stefano Borgani\inst{\ref{units},\ref{oats}},
    Francesco Calura\inst{\ref{oas}},
    Giulia Despali\inst{\ref{heidelberg}},
    Carlo Giocoli\inst{\ref{oas},\ref{infnbo},\ref{unibo}},
    Gian Luigi Granato\inst{\ref{oats},\ref{iateconicet},\ref{ifpu}},
    Claudio Grillo\inst{\ref{unimi},\ref{oami}},
    Lauro Moscardini\inst{\ref{unibo},\ref{oas},\ref{infnbo}}, 
    Elena Rasia\inst{\ref{oats}, \ref{ifpu}}, 
    Piero Rosati\inst{\ref{unife}},
    Giuseppe Angora\inst{\ref{unife},\ref{oana}},
    Luigi Bassini\inst{\ref{ctac}},
    Pietro Bergamini\inst{\ref{unimi},\ref{oas}},
    Gabriel B. Caminha\inst{\ref{mpa}},
    Giovanni Granata\inst{\ref{unimi}},
    Amata Mercurio\inst{\ref{oana}},
    Robert Benton Metcalf\inst{\ref{unibo},\ref{oas}},
    Priyamvada  Natarajan\inst{\ref{yale}}, 
    Mario Nonino\inst{\ref{oats}},
    Giada Venusta Pignataro\inst{\ref{unibo},\ref{ira}},
    Cinthia Ragone-Figueroa\inst{\ref{iateconicet},\ref{oaargentina},\ref{oats}},
    Eros Vanzella\inst{\ref{oas}},
    Ana Acebron\inst{\ref{unimi}},
    Klaus Dolag\inst{\ref{usm},\ref{mpa}},
    Giuseppe Murante\inst{\ref{oats}},  
    Giuliano Taffoni\inst{\ref{oats}},
    Luca Tornatore\inst{\ref{oats}}, 
    Luca Tortorelli\inst{\ref{usm}},
    Milena Valentini\inst{\ref{usm}}
    }
   \institute{
    INAF-Osservatorio di Astrofisica e Scienza dello Spazio di Bologna, Via Piero Gobetti 93/3, I-40129 Bologna, Italy, \label{oas}
    \and
    INFN-Sezione di Bologna, Viale Berti Pichat 6/2, I-40127 Bologna, Italy\label{infnbo}
    \and
    Dipartimento di Fisica e Astronomia "Augusto Righi", Alma Mater Studiorum Università di Bologna, via Gobetti 93/2, I-40129 Bologna, Italy\label{unibo}
    \and
    INAF - Osservatorio Astronomico di Trieste, via G.B. Tiepolo 11, I-34143 Trieste, Italy\label{oats}
    \and
    IFPU - Institute for Fundamental Physics of the Universe, Via Beirut 2, I-34014 Trieste, Italy\label{ifpu}
    \and
    Astronomy Unit, Department of Physics, University of Trieste, via Tiepolo 11, I-34131 Trieste, Italy \label{units}%\\
    \and Zentrum für Astronomie der Universität Heidelberg, Institut für Theoretische Astrophysik, Albert-Ueberle-Str. 2, D-69120 Heidelberg, Germany\label{heidelberg}
    \and
    Instituto de Astronom\'ia Te\'orica y Experimental (IATE), Consejo Nacional de Investigaciones Cient\'ificas y T\'ecnicas de la\\ Rep\'ublica Argentina (CONICET), Universidad Nacional de C\'ordoba, Laprida 854, X5000BGR, C\'ordoba, Argentina\label{iateconicet}
    \and Dipartimento di Fisica, Università degli Studi di Milano, Via Celoria 16, I-20133 Milano, Italy\label{unimi}
    \and INAF - IASF Milano, via A. Corti 12, I-20133 Milano, Italy\label{oami}
    \and 
    Dipartimento di Fisica e Scienze della Terra, Università degli Studi di Ferrara, Via Saragat 1, I-44122 Ferrara, Italy\label{unife}
    \and
    Center for Theoretical Astrophysics and Cosmology, Institute for Computational Science, University of Zurich, Winterthurerstrasse 190, CH-8057 Zürich, Switzerland\label{ctac}
    \and 
    INAF-Osservatorio Astronomico di Capodimonte, Via Moiariello 16, 80131 Napoli, Italy\label{oana}
    \and 
    Department of Astronomy, Yale University, New Haven, CT, USA\label{yale}
    \and
    Observatorio Astronómico de Córdoba, Universidad Nacional de Córdoba, Laprida 854, X5000BGR, Córdoba, Argentina\label{oaargentina}
    \and
    Universitäts-Sternwarte, Fakultät für Physik, Ludwig-Maximilians-Universität München, Scheinerstr.1, 81679 München, Germany \label{usm}%\\
    \and
    Max-Planck-Institut f\"{u}r  Astrophysik (MPA), Karl-Schwarzschild Strasse 1, D-85748 Garching bei M\"{u}nchen, Germany\label{mpa}%\\
    \and
    INAF - Istituto di Radioastronomia di Bologna, Via Gobetti 101, I-40129 Bologna, Italy\label{ira}
            }

   \date{Received April 15, 2022}

% \abstract{}{}{}{}{} 
% 5 {} token are mandatory
 
  \abstract
  % context heading (optional)
  % {} leave it empty if necessary  
   {\cite{2020Sci...369.1347M} recently reported an excess of galaxy-galaxy strong lensing (GGSL) in galaxy clusters compared to expectations from the $\Lambda$CDM cosmological model. Theoretical estimates of the GGSL probability are based on the analysis of numerical hydrodynamical simulations in the $\Lambda$CDM cosmology.}
  % aims heading (mandatory)
   {We quantify the impact of the numerical resolution and AGN feedback scheme adopted in cosmological simulations on the predicted GGSL probability, and determine if varying these simulation properties can alleviate the gap with observations.}
  % methods heading (mandatory)
   {We repeat the analysis of \cite{2020Sci...369.1347M} on cluster-size halos simulated with different mass and force resolutions and implementing several independent AGN feedback schemes.}
  % results heading (mandatory)
   {We find that improving the mass resolution by a factor of ten and twenty-five, while using the same galaxy formation model that includes AGN feedback, does not affect the GGSL probability. We find similar results regarding the choice of the gravitational softening. On the contrary, adopting an AGN feedback scheme that is less efficient at suppressing gas cooling and star formation leads to an increase in the GGSL probability by a factor between three and six. However, we notice that such simulations form overly massive subhalos whose contribution to the lensing cross-section would be significant while their Einstein radii are too large to be consistent with the observations. The primary contributors to the observed GGSL cross-sections are subhalos with smaller masses, that are compact enough to become critical for lensing. The population with these required characteristics appears to be absent in simulations.}
  % conclusions heading (optional), leave it empty if necessary 
   {Based on these results, we reaffirm the tension, previously reported in \cite{2020Sci...369.1347M}, between observations of GGSL and theoretical expectations in the framework of the $\Lambda$CDM cosmological model. The GGSL probability is sensitive to the galaxy formation model implemented in the simulations. Still, all the tested models have difficulty reproducing the stellar mass function and the internal structure of galaxies simultaneously.}


   \maketitle
%
%-------------------------------------------------------------------
\section{Introduction}
In the cold-dark-matter paradigm (CDM),  dark-matter halos form hierarchically, with the most massive systems resulting from mergers between smaller ones. Thus, dark matter halos contain a full hierarchy of substructures in the form of subhalos \citep{2008MNRAS.386.2135G,2010MNRAS.404..502G}. 

Galaxy clusters are ideal astrophysical laboratories to test this prediction of the CDM paradigm because their content is dark-matter-dominated. With virial masses as large as a few times $10^{15}\; M_\odot$, they are the strongest gravitational lenses in the universe. Strong gravitational lensing occurs when distant background galaxies are in near-perfect alignment with the massive foreground cluster, and the deflection of light by the gravity of the cluster results in highly distorted, multiple images of individual background galaxies. The extended dark matter distribution in cluster halos is responsible for most of these features. Galaxy clusters simultaneously split numerous distant sources into multiple images and produce highly distorted gravitational arcs over regions of size of the order of $\sim 1$ arcmin. At the same time, the larger scale dense cluster environments enhance the strong lensing effects produced on scales of a few arcseconds by the subhalos they host. Occasionally, arclets and additional multiple images of distant galaxies appear around individual cluster member galaxies. We can use these Galaxy-Galaxy Strong Lensing effects (GGSL hereafter) inside clusters  to infer the mass of the subhalos in which the cluster galaxies are embedded \citep[][]{1997MNRAS.287..833N,natarajan2002,2009ApJ...693..970N}.

Mass mapping via gravitational lensing has become an increasingly popular method to constrain the matter distribution in these objects \citep[see e.g.][]{1987A&A...172L..14S,FO88.1,LY89.1,1996ApJ...471..643K,BR05.1,2005MNRAS.359..417S,2007ApJ...668..643L,2007NJPh....9..447J,2008A&A...489...23L,2009MNRAS.395.1319J,2011A&ARv..19...47K,2010MNRAS.404..325R,2012ApJS..199...25P, 2012A&A...544A..71L, 2015MNRAS.447.1224M, 2016A&A...588A..99L,2017MNRAS.466.4094M}. The high resolution of the imaging cameras aboard the Hubble Space Telescope (HST) has greatly impacted our ability to identify multiply imaged galaxies in the fields of several galaxy clusters. Particularly significant are several multi-cycle HST programs which recently targeted strong lensing clusters, such as the {\em Cluster Lensing And Supernova survey with Hubble} \citep[CLASH,][]{2012ApJS..199...25P}, the {\em Hubble Frontier Fields} initiative \citep[HFF,][]{2017ApJ...837...97L}, the {\em  Reionization Lensing Cluster Survey} \citep[RELICS, ][]{2018ApJ...859..159C}, and the {\em Beyond Ultra-deep Frontier Fields And Legacy Observations}  \citep[BUFFALO,][]{2020ApJS..247...64S}. In these observational programs, several tens to hundreds of multiply imaged candidates, and strongly lensed galaxies have been identified. 

Follow-up spectroscopy by several independent groups has been ongoing for the bright, highly magnified multiple images in these clusters \citep{2014Msngr.158...48R,2013A&A...559L...9B,2015ApJ...812..114T,2017A&A...607A..93C,2021A&A...656A.147M} as well as for the more challenging fainter objects. Integral field observations with the  Multi-Unit Spectroscopic Explorer (MUSE) at the VLT have contributed invaluably to strong lensing modeling \citep[see, e.g.][]{2015A&A...574A..11K,2015ApJ...800...38G,2016ApJ...822...78G,2021A&A...646A..57V,2019A&A...632A..36C,2021MNRAS.508.1206J,2021A&A...645A.140B,2021A&A...646A..83R,2021arXiv210709079G}. Thanks to its tremendous efficiency and sensitivity to line emitters, this instrument has expanded the list of confirmed strongly lensed galaxies available for mapping the matter distribution and made unprecedented resolution possible in several galaxy clusters. Mass models have been built incorporating these constraints. They include self-similar smooth mass components to describe the large-scale cluster dark matter halos and small-scale dark matter substructures. These are generally assumed to be traced by bright cluster galaxies \citep{1997MNRAS.287..833N,2014MNRAS.444..268R,2015MNRAS.451.3920D,2015MNRAS.452.1437J,2015ApJ...814...69A,2015ApJ...800...38G,2016ApJ...817...60T,2016ApJ...819..114K,2016A&A...587A..80C,2016MNRAS.457.2029J,2016A&A...588A..99L,2016ApJ...822...78G,2016MNRAS.459.3447D,2017A&A...607A..93C,2017MNRAS.469.3946L,2017A&A...600A..90C}. VLT/MUSE spectroscopy has been critical to also produce complete samples of confirmed cluster members for this purpose. 

From the theoretical point of view, we can study the internal structure of galaxy clusters in the framework of CDM  with numerical hydrodynamical simulations. These simulations allow us to investigate the growth and the evolution of galaxy clusters in the cosmological context. If their resolution is high enough, we can use them to predict several properties of the subhalos, such as their mass profiles, abundance, radial distribution function, and even their capacity to produce strong lensing effects \citep{2017MNRAS.469.1997D,2020MNRAS.491.1295D,2022MNRAS.510.2480D}. 

Some studies in the literature already use the outcome of strong lensing models and hydrodynamical simulations to test these predictions of the CDM paradigm \citep[e.g.][]{PNSpringel2004}. Simulations show that the mass and radial distributions of subhalos are nearly universal, with the former being a (truncated) power-law with slope $\alpha \approx -0.9$ \citep[see, e.g.][]{2000ApJ...544..616G,PNdeLucia2007,2009MNRAS.399..497D,2010MNRAS.404..502G,2017MNRAS.470.4186Bahe,2017MNRAS.469.1997D}, and the second being much less concentrated than that of the dark matter particles \citep{2004MNRAS.355..819G,2008MNRAS.391.1685S,2012MNRAS.425.2169G}. \cite{2015ApJ...800...38G} found that the HFF cluster MACSJ 0416.1-2403 contains an unexpectedly high number of large-mass subhalos (i.e. subhalos with circular velocities $v_c>100$ km/s) compared to simulations \citep[see also][]{2016ApJ...827L...5M,2018ApJ...864...98B}, while \cite{2017MNRAS.468.1962N} noticed that the radial distribution of the observed visible substructures in clusters Abell 2744, MACSJ 0416.1-2403, and  MACSJ 1149.5+2223 is inconsistent with numerical simulations. 

More recently, \cite{2020Sci...369.1347M} (hereafter M20) reported an excess of GGSL events in a sample of 11 observed galaxy cluster lenses compared to expectations from $\Lambda$CDM hydrodynamic simulations. This result suggests that observed cluster member galaxies are more compact than their simulated counterparts. This and the former discrepancies between simulations and observations may signal a potential problem with the CDM paradigm and/or yet undiagnosed systematic issues with simulations. In particular, subhalo properties may be affected by several physical processes which are known to be challenging to model and include realistically in the simulations. For example, numerical simulations show that satellite halos can be gradually destroyed after being accreted, until they are completely dissolved in the host halo  \cite[see, e.g.][]{2004MNRAS.352..535D,2005MNRAS.359.1029V,2008MNRAS.386.2135G,2010MNRAS.404..502G,2016MNRAS.457.1208H,2016MNRAS.458.2848J,2017MNRAS.468..885V,2019MNRAS.485.2287B}. While there are known physical mechanisms that lead to the disruption of the satellite halos, additional numerical effects may also be at play. Tidal stripping and tidal heating can indeed remove large fractions of the mass of dark matter subhalos. Baryonic physics and processes like cooling or energy feedback could impact the efficiency with which subhalos can be destroyed and also alter their internal structure \citep{1988ApJ...327..507F,2017MNRAS.471.1709G}. On the other hand, it is well known that subhalos can also dissolve in N-body simulations because of numerical artifacts - limited mass and force resolution in a simulation, a problem well known to computational cosmologists as {\em overmerging} \citep{1994ApJ...433..468C,1996ApJ...457..455M,1999ApJ...516..530K,2000astro.ph..8453V,2018MNRAS.474.3043V}. 

In this paper, we delve deeply into how the resolution and adopted AGN feedback schemes impact the GGSL probability measured in numerical hydrodynamical simulations of cluster assembly. To assess the relevance of these simulation properties, we repeat the analysis of M20 on a sample of galaxy clusters re-simulated with different particle masses, softening lengths, and AGN feedback schemes. We also compare the simulations to a selection of four observed galaxy clusters for which we recently constructed detailed strong lensing mass models.

The paper is structured as follows: in Sect.~\ref{sect:GGSL}, we introduce the method to compute the GGSL probability; in Sect.~\ref{sect:observations}, we present the observational data set; in Sect.~\ref{sect:simulations}, we introduce the simulation data sets discussing the differences between them. In Sect.~\ref{sect:results}, we show the results of the analysis performed on the different simulated data sets. We quantify the GGSL probability and determine which subhalos contribute primarily to the GGSL cross-section. Finally, we draw our conclusions in Sect.~\ref{sect:conclusions}.

%--------------------------------------------------------------------
\section{Probability of GGSL events}
\label{sect:GGSL}
To describe their strong lensing properties, we assume that the mass distribution of galaxy clusters is characterized by the projected surface mass density $\Sigma(\vec \theta)$.  The vector $\vec \theta$ indicates the angular position on the lens plane.

The lens convergence, $\kappa(\vec\theta)$, is defined as the ratio between the surface density and the {\em critical} surface density,
\begin{equation}
    \kappa(\vec\theta)=\frac{\Sigma(\vec\theta)}{\Sigma_{\rm cr}}\;,
\end{equation}
where
\begin{equation}
    \Sigma_{\rm cr}(z_l,z_s) \equiv \frac{c^2}{4\pi G}\frac{D_s}{D_{ls}D_l} \;.
\end{equation}
The quantities $D_l$, $D_{s}$, and $D_{ls}$ are the angular diameter distances between the observer and the lens, the observer and the source, and the lens and the source, respectively. Further details on lensing basics can be found in, e.g., \cite{Meneghetti2021} \citep[see also][for some reviews on cluster lensing]{2011A&ARv..19...47K,2013SSRv..177...31M,2020A&ARv..28....7U}.

A light ray emitted by a source at redshift $z_s$, crossing the lens plane at position $\vec\theta$, is deflected by the {\em reduced} deflection angle
\begin{equation}
    \vec\alpha(\vec\theta) =\frac{1}{\pi}\int \kappa(\vec\theta')\frac{\vec\theta-\vec\theta'}{|\vec\theta-\vec\theta'|^2}d^2\theta' \;,
    \label{eq:alpha}
\end{equation}
where the integral is extended to the whole lens plane. 

Using the first partial derivatives of the deflection angle components $\alpha_1$ and $\alpha_2$, we define the shear tensor, whose components $\gamma_1$ and $\gamma_2$ are
\begin{eqnarray}
   \gamma_1(\vec\theta) & = & \frac{1}{2}\left[\frac{\partial \alpha_1(\vec\theta)}{\partial \theta_1}-\frac{\partial \alpha_2(\vec\theta)}{\partial \theta_2}\right] \nonumber \\
   \gamma_2(\vec\theta) & = & \frac{\partial \alpha_1(\vec\theta)}{\partial \theta_2} = \frac{\partial \alpha_2(\vec\theta)}{\partial \theta_1} \;.
\end{eqnarray}
The shear modulus is $\gamma(\vec\theta)=\sqrt{\gamma_1^2(\vec\theta)+\gamma_2^2(\vec \theta)}$.

Following the definition of M20, the GGSL cross-section for a given source redshift, $z_s$, is the area on the source plane enclosed by cluster galaxies' tangential caustics. We compute it as follows:
\begin{itemize}
\item we begin from the map of the tangential eigenvalue of the lensing Jacobian, $\lambda_t(\vec \theta)$. This is defined as 
\begin{equation}
    \lambda_t(\vec \theta) = 1 -\kappa(\vec \theta)-\gamma(\vec \theta) \;;
\end{equation}
\item we find the tangential critical lines, which are the zero-level contours of $\lambda_t(\vec \theta)$:
\begin{equation}
    \vec \theta_t : \lambda_t(\vec\theta_t)=0 \; ;
\end{equation}
\item because of its multiple mass components, a galaxy cluster typically has many tangential critical lines.  The largest ones are associated with the overall smooth cluster dark matter halo.  We call these critical lines {\em primary}.  A cluster can have more than one primary critical line.  For example, a cluster can have multiple large-scale mass components, remnants from major mergers.  Each of these mass components has its critical line.  Our criterion to identify the primary critical lines is based on the size of the effective Einstein radius.  Given a critical line enclosing the area $A_c$, the Einstein radius is given by:
\begin{equation}
\theta_E = \sqrt{\frac{A_c}{\pi}} \;.
\label{eq:einstrad}
\end{equation}
We call {\em primary} those critical lines which have $\theta_E>5''$, independently of the source redshift;
\item on the contrary, {\em secondary} tangential critical lines associated with the galaxy-scale subhalos have smaller Einstein radii.  In particular, we consider those critical lines satisfying the condition $0.5''\leq \theta_E \leq 3''$, regardless of the source redshift. The lower limit on the Einstein radius is motivated by the fact that smaller critical lines may not be properly resolved in the numerical simulations that we consider in this paper.  We decide to exclude the secondary critical lines with $\theta_E>3''$ for reasons that will be explained in Sect.~\ref{sect:substhetae};
\item once the critical lines have been identified, we map them onto the source plane using the lens equation \citep[see, e.g.,][]{Meneghetti2021}, and we obtain the corresponding caustics:
\begin{equation}
    \vec\beta_c = \vec\theta_c-\vec\alpha(\vec\theta_c) \;.
\end{equation}
\item for each caustic, we measure the enclosed area, $\sigma_i$.  Summing the areas of all caustics, we obtain the total cluster GGSL cross-section, $\sigma_{\rm GGSL}(z_s)=\sum{\sigma_i(z_s)}$. We can repeat this procedure by changing $z_s$ to measure how the cross-section varies as a function of the source redshift.
\end{itemize}

To compute the probability of GGSL events for sources at a given redshift, we divide the total cross-section by the area in the source plane, which corresponds, via the lens equation, to the region relevant for the lens deflection angle field, $A_s(z_s)$:
\begin{equation}
    P_{\rm GGSL}(z_s)=\frac{\sigma_{\rm GGSL}(z_s)}{A_s(z_s)} \;.
\end{equation}

Note that we compute the cross-sections in the limit of point sources. The cross-sections would be larger for extended sources. Indeed, in this case strong lensing effects can occur even if the sources are not entirely contained by the caustics. Thus, the values reported in this paper for both observed and simulated lenses are lower limits of the true GGSL cross-sections and probabilities.

\section{Application to observations}
\label{sect:observations}
In this paper, we focus on four galaxy clusters, namely Abell S1063 \citep[$z=0.3457$,][]{2013A&A...559L...9B}, MACS J0416.1-2403 \citep[$z=0.397$,][]{2016ApJS..224...33B}, MACS J1206.2-0847 \citep[$z=0.439$,][]{2013A&A...558A...1B}, and PSZ1 G311.65-18.48 \cite[$z=0.443$,][]{2016A&A...590L...4D}. These clusters are well known strong gravitational lenses. Deep multi-band observations with Hubble have revealed several tens of background galaxies that appear as distorted arcs and sets of multiple images \citep{2012ApJS..199...25P,2017ApJ...837...97L,2019Sci...366..738R,2021A&A...655A..81P}. The HST images also contain several examples of GGSL events \citep{2014ApJ...786...11G,2016MNRAS.458.1493P,2017A&A...607A..93C,2017ApJ...842...47V,2018MNRAS.479.2630D,2020Sci...369.1347M}.

By modeling these large sets of strong lensing features, we have resolved the inner structure of the dark matter halos in these clusters down to galaxy scales. The details of the reconstructions themselves can be found in a series of prior papers \citep{2019A&A...631A.130B,2021A&A...645A.140B,2021A&A...655A..81P}.
In short, we use the so-called parametric approach implemented in the publicly available lens inversion code {\sc Lenstool} \citep[see e.g.][]{1996ApJ...471..643K,2007NJPh....9..447J,2011A&ARv..19...47K,2017MNRAS.472.3177M}. The adopted fiducial model characterizes the cluster as a combination of mass components whose density profiles are analytic functions dependent on a few parameters. In addition, it assumes that luminous cluster galaxies trace the mass in the dark matter substructure. The contribution of the entire population of cluster members to the cluster mass budget is modeled using scaling relations such as the Faber-Jackson \citep{1976ApJ...204..668F,1997MNRAS.287..833N}. We use MUSE integral field spectroscopy to measure the internal kinematics of a subset of cluster galaxies and use them to calibrate these scaling relations.
In lens optimization, one searches for the best-fit parameters that minimize the distance between the observed and model-predicted multiple images. The optimization result is a projected mass map fully described by parameters that allow us to disentangle the mass into the large-scale cluster dark matter halo and the smaller-scale galaxy components. 

\begin{figure*}
   \centering
   \includegraphics[width=17cm]{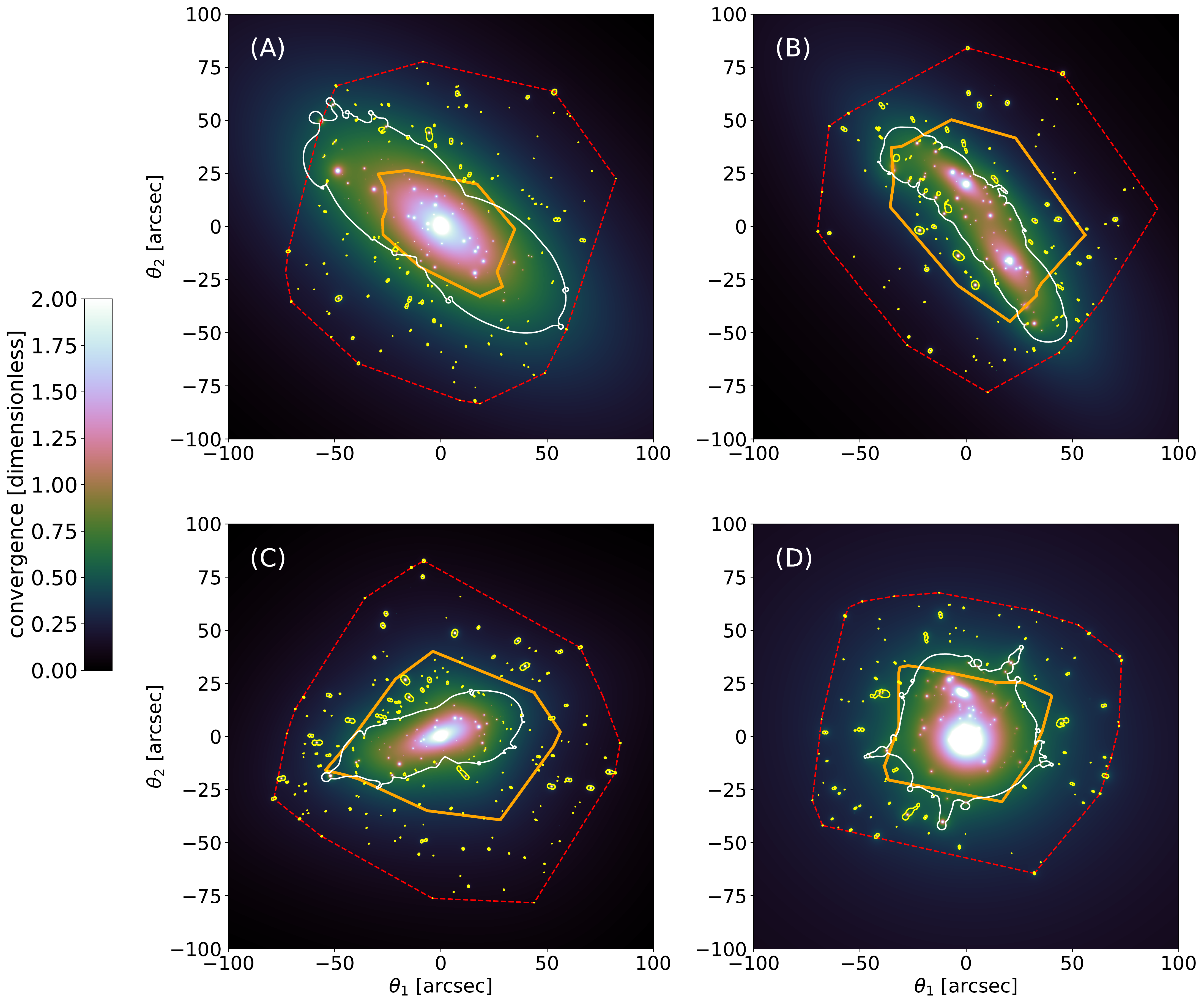}
      \caption{Reconstructed convergence maps for $z_s=3$ of Abell S1063 (Panel A), MACS J0416.1-2403 (Panel B), MACS J1206.2-0847 (Panel C), and PSZ1 G311.65-18.48 (Panel D). We also show the primary and secondary critical lines (white and yellow solid lines, respectively). The red dashed lines mark the region in the lens plane containing all the cluster members included in the mass models. When mapped onto the source plane, these lines correspond to the thick solid orange lines, which show the size of the region in the source plane containing all the secondary caustics.}
         \label{fig:clusterreco}
\end{figure*}

In Fig.~\ref{fig:clusterreco}, we show the convergence maps of the four clusters with overlaid the primary and secondary tangential critical lines for $z_s=3$ (white and yellow solid lines, respectively). We use these models to compute the GGSL cross-sections and probabilities as outlined in Sect.~\ref{sect:GGSL}.

The selection of the cluster members included in the lens models is relevant for this analysis. Only cluster galaxies within the red dashed lines in Fig.~\ref{fig:clusterreco} contribute to the cluster deflection fields. We identified them either spectroscopically or because their colors are consistent with the cluster red sequences. In addition, we retained only galaxies with apparent magnitude $m_{\rm F160W} \leq 24$ in the HST/WFC3 F160W band. Only the galaxies included in the lens model are assumed to contribute to the GGSL cross-section.

Most of the cluster galaxies are massive and compact enough to have their own (secondary) critical lines. Because of the cluster magnification, the corresponding caustics occupy regions in the source plane significantly smaller than those enclosed by the dashed red lines. We mark these regions in Fig.~\ref{fig:clusterreco} with thick solid orange lines. To calculate the GGSL probability, we divide the GGSL cross-sections by the areas $A_s$ enclosed within these lines. This approach is  different from that utilized in M20, who mapped the entire field-of-view of $200''\times200''$ onto the source plane to compute $A_s(z_s)$. This field-of-view, however, is larger than the reconstructed cluster region. A significant portion of it does not contain cluster galaxies because they were not included in the lens model. For this reason, the GGSL probabilities calculated in M20 for Abell S1063, MACS J0416.1-2403, and MACS J1206.2-0847 (these clusters were part of the 11 cluster sample used in the M20 analysis) are smaller than those reported here in this paper.

\section{Application to numerical simulations}
\label{sect:simulations}

\subsection{Numerical data sets}
The simulated cluster halos used in this paper belong to a suite of numerical hydrodynamical simulations, dubbed the {\em Dianoga} suite, that have been extensively studied in several previous works, including several lensing analyses \citep[e.g.][]{2010A&A...514A..93M,2012MNRAS.427..533K,2012NJPh...14e5018R}. We focus on a sample of seven cluster-size halos. They were first identified in a low--resolution periodic simulation box with co-moving size of $\sim 1.4$ Gpc for a flat $\Lambda$CDM model with present matter density parameter $\Omega_{m,0}=0.24$ and baryon density parameter $\Omega_{b,0}=0.04$. The Hubble constant adopted was $H_0=72$ km $s^{-1}$ Mpc$^{-1}$ and the normalisation of the matter power spectrum on a scale of $8\,h^{-1}$ Mpc was $\sigma_8=0.8$, where $h=H_0/100$. The primordial power spectrum of the density fluctuations adopted was $P(k) \propto k^{n}$ with $n=0.96$. The parent simulation followed 1024$^{3}$ collision-less particles in the box. The clusters were identified at $z=0$ using a standard {\it Friends-of-Friends} algorithm, and their Lagrangian region was then re-simulated at higher resolution employing the {\it Zoomed Initial Conditions} code \citep[ZIC;][]{1997MNRAS.286..865T,2011MNRAS.418.2234B}. The resolution is progressively degraded outside this region to save computational time while still ensuring that the larger scale tidal field is accurately described. The Lagrangian region was large enough to ensure that only high-resolution particles were present within five virial radii of the clusters.

The re-simulations were then carried out using the TreePM--SPH {\small GADGET--3} code \citep{SP05.1}, adopting an improved Smoothed Particle Hydrodynamics (SPH) solver \citep{2016MNRAS.455.2110B}.
Our treatment implements several recipes for the relevant physical processes that operate, and these are  summarized as follows. Metallicity-dependent radiative cooling and the effect of a uniform time-dependent UV background are modeled as in \cite{2014MNRAS.438..195P}. A sub-resolution model for star formation from a multi-phase interstellar medium is implemented as in \cite{2003MNRAS.339..289S}. Kinetic feedback driven by a supernova (SN) is in the form of galactic winds. Metal production from SN-II, SN-Ia, and asymptotic-giant-branch stars follows the recipe by \cite{2007MNRAS.382.1050Tornatore}. 

\begin{figure*}
   \centering
   \includegraphics[width=1.0\linewidth]{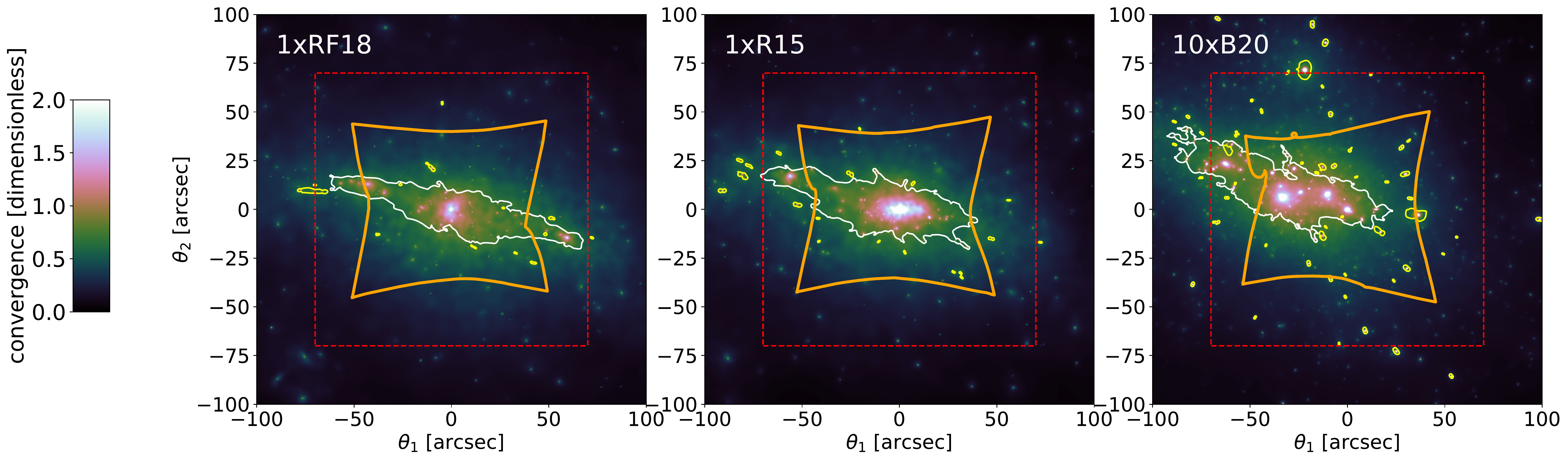}
      \caption{Examples of convergence map for $z_s=3$ for a simulated cluster at $z_l \sim 0.4$. The left, central, and right panels show the maps for the same halo in the 1xRF18, 1xR15, and 10xB20 samples. The solid white and yellow lines show the primary and secondary tangential critical lines. The dashed red line indicates the region where we identify the secondary critical lines for measuring the GGSL probability. The solid orange line shows the boundaries of this region on the source plane.}
\label{fig:examplesim}
\end{figure*}

The seven cluster halos that we study in this paper were simulated with different mass and spatial resolutions and implemented galaxy formation models characterized by various independent AGN-feedback schemes. In all cases, the simulations began from the same initial conditions. As specified below, the details of these models can be found elsewhere, here we briefly summarize them.

\paragraph{1xRF18:} The first set of simulations, which we dub 1xRF18, implements the feedback scheme proposed by \cite{2013MNRAS.436.1750R} \citep[based on previous work by ][]{2005MNRAS.361..776Springel} with some modifications as outlined in \cite{2018MNRAS.479.1125R}. 
The gas particles are assumed to be multi-phase when their density exceeds the threshold of $0.1$ cm$^{-3}$, and their temperature is $T<2.5\times 10^{5}\,K$. Multi-phase particles comprise a cold and a hot phase in pressure equilibrium. Particles in the cold phase can cool and form stars.
Black Holes (BHs) with an initial mass of $M_{\rm BH}\sim 7\times 10^6\,M_\odot$ are seeded in subhalos with mass $M>3\times 10^{11}\,M_\odot$. The Eddington limited gas accretion rate onto BHs, $\dot{M}_{\rm BH}$, is computed by multiplying the Bondi rate by a boost factor $\alpha$. A distinction is made between the cold and hot accretion modes. The gas temperature threshold separating these two accretion modes is set at $T=5\times 10^5\,K$. Cold and hot accretion correspond to two different boost factors, namely $\alpha_{cold}=100$ and $\alpha_{hot}=10$, motivated to match observational constraints at $z = 0$. The rate of available energy feedback from BH accretion is $\dot{E}=\epsilon_f\epsilon_r\dot{M}_{\rm BH}c^2$, where the parameters $\epsilon_r$ and $\epsilon_f$ describe the fraction of accreted mass transformed into radiation and the fraction of radiated energy thermally coupled to the gas particles, respectively. The scheme also accounts for cold cloud evaporation. The parameters $\epsilon_r$ and $\epsilon_f$ are calibrated to reproduce the observed relation between BH mass and stellar mass in spheroids, i.e., the Magorrian relation \citep{1998AJ....115.2285M}. Specifically, the 1xRF18 simulations adopt $\epsilon_r=0.07$ and $\epsilon_f=0.1$ and assume a transition from a quasar mode to a radio mode AGN feedback when the accretion rate becomes smaller than a given fraction of the Eddington limit, $\dot{M}_{\rm BH}/\dot{M}_{\rm Edd}=10^{-2}$. In radio mode, the feedback efficiency $\epsilon_f$ is increased to $0.7$. 

To counteract numerical effects that tend to move BHs away from the stellar system in which they were first seeded, \cite{2018MNRAS.479.1125R} also implement an algorithm to keep the BHs at the center of their DM halos. Pinning the BHs is particularly relevant for this study because the absence of AGN feedback at the center of massive galaxies due to BH wandering leads to catastrophic cooling and excessive star formation that may artificially increase the GGSL cross-section. \cite{2018MNRAS.479.1125R} show that the energy feedback model implemented in their simulations, which comprise the seven clusters in our sample, produce Brightest Cluster Galaxies (BCGs) with stellar masses in excellent agreement with the observations. \cite{2019A&A...630A.144B} also show that these simulations predict a stellar mass function that matches the observations at the high-mass end but is lower by a factor of $\sim 2$ at masses lower than $8.5\times 10^{11}\,M_\odot$. 

The mass resolution for the DM and gas particles is $m_{DM} = 1.1 \times 10^9 \, M_\odot$ and $m_{gas} = 2.1 \times 10^8 \,M_\odot$, respectively. For the gravitational force, a Plummer-equivalent softening length of $\epsilon  = 7.7$ ckpc (comoving kpc) is used for DM and gas particles, whereas $\epsilon = 4.2$ ckpc for BH and star particles. 

\paragraph{1xR15:} The 1xR15 sample is a sub-sample of the simulations described in \cite{2015ApJ...813L..17R}. Some of these clusters were also analyzed in M20. The AGN feedback model implemented in these simulations is presented in detail in \cite{2015MNRAS.448.1504Steinborn} and \cite{2014MNRAS.442.2304Hirschmann}. The main difference with the model used in the 1xRF18 simulations is that the released thermal energy accounts for contributions by mechanical outflows and radiation, that are separately computed in the code. The outflow component dominates at accretion rates below $\sim 0.01 \dot{M}_{\rm Edd}$. This results in an additional parameter that describes the outflow efficiency $\epsilon_o$. The parameters $\epsilon_r$ and $\epsilon_o$ are not constant but are allowed to vary as a function of the BH mass and accretion rate. This implies a continuous transition between the feedback processes acting in the radio and quasar modes. The model is calibrated to reproduce the Magorrian relation ($\epsilon_f=0.05$), but it also reproduces quite well the observed stellar mass function over a wide range of masses \citep{2015MNRAS.448.1504Steinborn}.
The mass resolution is identical to the 1xRF18 simulations, but the Plummer-equivalent softening length is smaller: $5.2$ ckpc for the DM and gas particles and $2.7$ ckpc for BH and star particles.

\paragraph{10xB20:} The 10xB20 clusters are a sub-sample of the simulations presented by \cite{2020A&A...642A..37Bassini}. The mass resolution is a factor of 10 higher than in the 1xR15 and 1xRF18 simulations. Thus, the particle mass for the DM and gas particles is $m_{DM} = 1.1 \times 10^8 \, M_\odot$ and $m_{gas} = 2.1 \times 10^7 \,M_\odot$, respectively. The gravitational softening is $\epsilon=1.9$ ckpc for the DM and gas particles and $0.5$ ckpc for the BH and star particles. BHs with an initial mass of $M_{\rm BH}\sim 5.5\times 10^5\,M_\odot$ are seeded in subhalos whenever the following conditions are simultaneously fulfilled: (i) the total stellar mass is higher than $2.8\times 10^{9}\,M_\odot$; (ii) the stellar-to-DM mass ratio is higher than 0.05; (iii) the gas mass is equal to or larger than 10\% of the stellar mass; and (iv) no other central BH is already present.

\cite{2020A&A...642A..37Bassini} implement a feedback scheme similar to that of \cite{2018MNRAS.479.1125R}, but with few significant modifications. They do not impose a temperature threshold to define multi-phase gas particles, and the energy released by AGN feedback is not used to evaporate the cold phase of gas particles. Under these conditions, cold particles can cool more efficiently and form more stars. Thus, this model agrees better with the observed stellar mass function at intermediate masses, but it over-predicts the number of massive galaxies. As shown by \cite{2020A&A...642A..37Bassini}, these simulations also tend to produce overly massive BCGs compared to observations. This problem is common to several other simulation suites as reported in the literature, including the {\sc Hydrangea/Eagle}, Illustris-TNG, and {\sc FABLE} simulations \citep{2017MNRAS.470.4186Bahe,2018MNRAS.473.4077Pillepich,2020MNRAS.498.2114H}. \cite{2020A&A...642A..37Bassini} also implement a different algorithm to prevent BH wandering compared to \cite{2018MNRAS.479.1125R}. The parameters $\epsilon_r$ and $\epsilon_f$ are set to 0.07 and 0.16, respectively, to reproduce the Magorrian relation at $z = 0$. As we discuss later in this paper, we simulate two subsets of the Dianoga cluster sample with the same AGN feedback model implemented in the 10xB20 simulations, but with mass resolution lower by a factor of 10 and higher by a factor of 2.5, respectively.   

\subsection{Lensing analysis}

For each of the seven cluster halos in all simulation sets, we consider the mass distributions at six different redshifts between $z_{min}=0.24$ and $z_{max}=0.55$. The four clusters in the observational data set have redshifts in this range. Thus, we can compare them safely to the simulations.

\begin{figure*}
   \centering
   \includegraphics[width=1.0\linewidth]{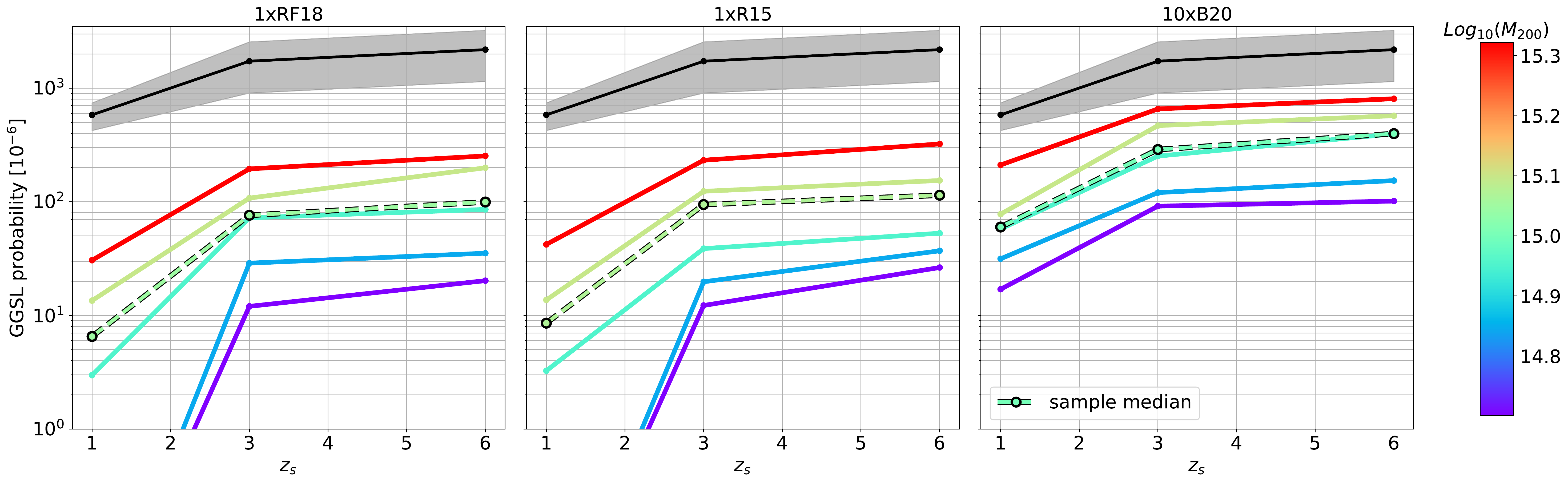}
      \caption{The GGSL probability as a function of the source redshift. The mean GGSL
probability among our observational sample is shown with the solid black line in all panels. The gray color band shows the 100\% confidence interval. The results for the 1xRF18, 1xR15, and 10xB20 simulation sets are shown in the left, central, and right panels, respectively. Each colored solid line corresponds to the median of GGSL probability in mass bins. The colors reflect the cluster mass, as indicated in the color bar on the right. The dashed black lines show the median probability among all cluster projections.}
\label{fig:ggsl_prob}
\end{figure*}

For each simulation snapshot, we generate three lens planes by projecting the particles within cylinders of depth $10$ Mpc along the axes of the simulation box. We follow the procedure outlined in M20 to produce the surface density maps. We use the python code {\sc Py-SPHviewer} \citep{alejandro_benitez_llambay_2015_21703} to smooth the particle mass distributions using an adaptive Smoothed-Particle-Hydrodynamics (SPH) scheme. We convert the surface density maps into deflection angle maps using Fast-Fourier-Transform techniques to solve Eq.~\ref{eq:alpha} and compute the GGSL cross-sections and probabilities as explained in Sect.~\ref{sect:GGSL}. M20 measured the GGSL probability in regions of $200\times 200$ arcsec centered on each cluster. This paper considers smaller areas of $150''\times150''$, comparable to those within which we identified the cluster members included in the lens models in the observational data set. These regions correspond to smaller $A_s$ when mapped onto the source planes (as shown by the solid orange line in Fig.~\ref{fig:examplesim}). As a result, the GGSL probabilities reported in this paper are higher than those quoted by M20 by a factor $\sim 1.5$. We adopted this new area constraint to better match the various simulated sub-samples studied here.

For each simulation set, our sample consists of 126 lens planes. We discard those corresponding to cluster masses $M_{200}<5\times 10^{14}\;M_\odot$, where $M_{200}$ is the mass within the radius enclosing a mean density of $200\times \rho_{\rm crit}$, and $\rho_{\rm crit}$ is the critical density of the universe. After this selection, the median $M_{200}$ masses of the 1xRF18, 1xR15, and 10xB20 samples are $1.34\times10^{15}\,M_\odot$, $1.32\times10^{15}\,M_\odot$, and $1.1\times10^{15}\,M_\odot$, respectively. Based on weak and strong lensing analyses, Abell S1063, MACS J0416.1-2403, MACS J1206.2-0847 have estimated masses in the range $10^{15} \lesssim M_{200} \lesssim 2\times 10^{15}\;M_\odot$ \citep{2014ApJ...795..163U,2015ApJ...806....4M,2016ApJ...821..116U,2018ApJ...860..104U}. The mass of PSZ1 G311.65-18.48 estimated from {\em Planck} SZ data is $M_{500}\sim 6.6 \times 10^{14}$ \citep{2014A&A...571A..20P,2016A&A...590L...4D}, where $M_{500}$ is the mass corresponding to an overdensity of $500\times \rho_{crit}$. Converting to $M_{200}$ assuming a typical concentration-mass relation \citep{hu2003}, we obtain $M_{200}\sim 8.5\times 10^{14}\,M_\odot$. 

We show examples of convergence maps for the same cluster halo in the three simulation sets in Fig.~\ref{fig:examplesim}. The source redshift is $z_s=3$. We notice that the halo in the 10xB20 simulation contains several massive subhalos that are not equally prominent in the 1xR15 and 1xRF18 maps. As in Fig.~\ref{fig:clusterreco}, the white and yellow solid lines indicate the primary and secondary tangential critical lines, respectively.

\section{Results}
\label{sect:results}
\subsection{GGSL probabilities}

The black solid lines in the three panels of Fig.\ref{fig:ggsl_prob} show the mean GGSL probability in the observational data set as a function of the source redshift. The gray colored band corresponds to the 99.9\% confidence interval. We show the results for the simulated clusters using colored solid lines. The colors indicate the cluster mass, $M_{200}$. We group the cluster projections into 5 mass bins, whose edges are $5\times10^{14}\,M_\odot$, $6\times10^{14}\,M_\odot$, $8\times10^{14}\,M_\odot$, $10^{15}\,M_\odot$, $1.5\times10^{15}\,M_\odot$, and $3\times10^{15}\,M_\odot$. Although the data sets contain only seven clusters, we remind that we project each of them along three orthogonal lines of sight. In addition, we use snapshots of each cluster at six different redshifts. 
The left, central, and right panels refer to the 1xRF18, 1xR15, and 10xB20 simulation sets, respectively. The dashed lines show the medians among all cluster projections in each data set and are colored according to the median mass in each sample. 

All simulated clusters, independent of the simulation set they belong to, have GGSL probabilities consistently smaller than those of the observational data set. For the samples 1xRF18 and 1xR15, the results are very similar. The cluster halos with the highest GGSL probabilities fall short of the observations by a factor $\sim 7$ at $z_s>3$, on average. The median GGSL probability in these simulations is lower by more than one order of magnitude than in the observational set, thus confirming the results of M20.

The efficiency of the AGN feedback models implemented in these two types of simulations is similar (as discussed in \cite{ragagnin22}). The softening length of the 1xRF18 simulations is larger than that of the 1xR15 simulations, the cluster halos in 1xRF18 sample contain $\sim 25\%$ fewer low-mass subhalos ($M_{sub}\lesssim 10^{11}\, M_\odot$) in their inner regions. Interestingly, the nearly identical GGSL probabilities we measure in the 1xRF18 and 1xR15 simulations indicate that softening has a low impact on the GGSL cross-sections of the simulated clusters. Thus, most of the GGSL signal in these halos originates from subhalos of mass $M_{sub}\gtrsim 10^{11}\,M_\odot$. On the contrary, as reported in Fig.~S5 of M20, the GGSL events observed in MACS J0416.1-2403 and MACS J1206.2-0847 are produced by cluster galaxies with estimated masses $M_{sub}<10^{11}\,M_\odot$.

As shown in the right panel of Fig.~\ref{fig:ggsl_prob}, the gap between the 10xB20 simulations and the observational data sets is significantly smaller. For example, the simulated halos with the highest GGSL probabilities fall short of the observations by only a factor of three in this case. Still, however, the median GGSL probability is smaller than the observed value by a factor of five. 

\begin{figure}
   \centering
   \includegraphics[width=9cm]{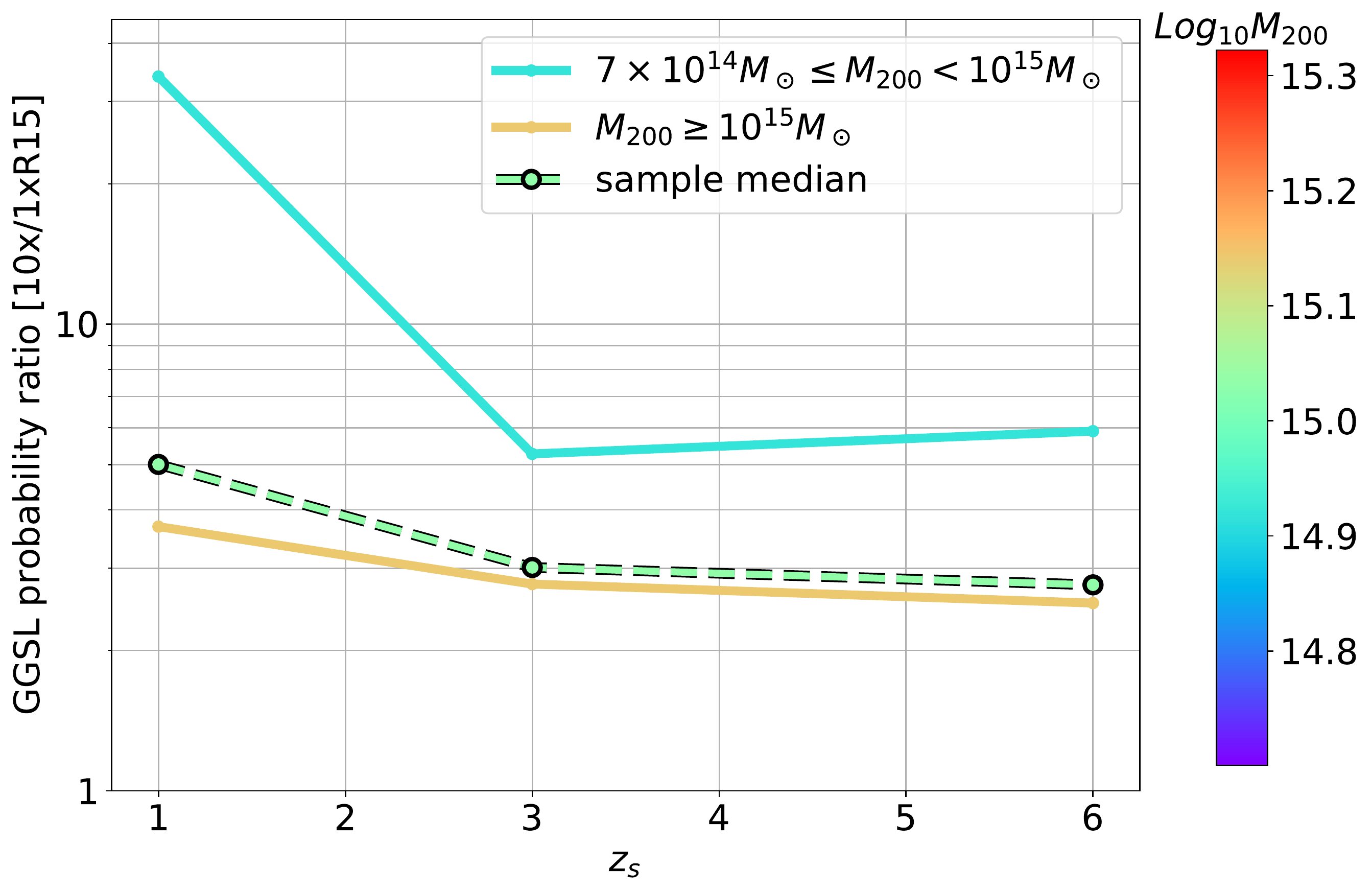}
      \caption{The median ratio of the GGSL probabilities in the 10xB20 sample to those in the 1xR15 sample as a function of the source redshift. We show the results separately for clusters with masses larger and smaller than $10^{15} \,M_\odot$, as shown by the different colors of the solid lines. The dashed line shows the median ratio of the GGSL probabilities in the whole sample.}
\label{fig:mass_dep}
\end{figure}

The increment of GGSL probability in 10xB20 compared to the 1xRF18 and 1xR15 simulations depends on cluster mass. As shown in Fig.~\ref{fig:mass_dep}, it is larger for low-mass than for high-mass clusters. If we consider cluster halos in two mass bins, corresponding to masses  $7\times 10^{14}\,M_\odot \leq M_{200} < 10^{15}\,M_\odot$ and $M_{200} \geq 10^{15}\,M_\odot$ we can compute the median ratios of the GGSL probabilities of the 10xB20 clusters to those of the 1xR15 in the two bins. For the most massive halos, the ratio is $\lesssim 3$ at $z_s \geq 3$, while it is $\lesssim 6$ for the clusters in the smallest mass bin. As explained above, such an increment is still insufficient to fill the gap with the observations.   

All the simulated cluster samples contain a majority of clusters with masses $M_{200}>10^{15}\,M_\odot$. Consequently, the median GGSL probability increment for the whole sample is similar to that of the most massive clusters, except at $z_s=1$. For low source redshifts, the differences between 10xB20 and 1xR15 simulations in the smallest mass bin are much more significant and amount to a factor of $\sim 35$.

Subhalos in low mass clusters in the 1x simulations are often sub-critical for lensing, i.e., they do not develop secondary critical lines, especially at low source redshifts. Therefore it is not surprising that the differences between the 10xB20 and 1x simulations emerge more significantly at small cluster masses and for low $z_s$.

\subsection{Effects of resolution and AGN feedback}

\begin{figure*}
   \centering
   \includegraphics[width=1.0\linewidth]{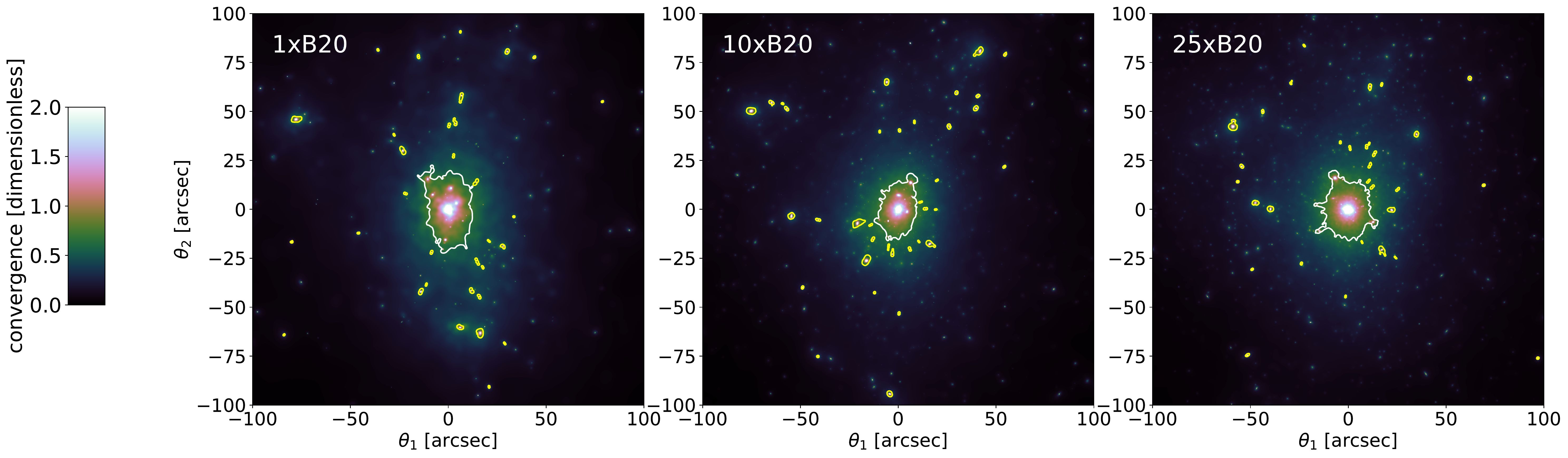}
      \caption{Convegence maps for $z_s=3$ of a cluster halo simulated with the same AGN feedback scheme of the 10xB20 simulations, but with different mass resolutions and softening lengths. See text for more details.}
\label{fig:examples_sim_feedback_res}
\end{figure*}

We want to establish if the higher GGSL probability in the 10xB20 sample compared to the 1x samples is mainly due to these simulations' higher spatial and force resolutions or to their less efficient feedback scheme. In higher resolution simulations, the inner structure of subhalos is better resolved, preventing the smaller subhalos from being destroyed due to numerical effects. At the other end, a lower feedback efficiency implies more substantial cooling and higher star formation that may cause the formation of denser subhalos that are more powerful strong lenses.

The Dianoga suite contains simulations (although for only four cluster halos) carried out at the exact mass resolution of the 1xR15 and 1xRF18 samples, but with the same AGN feedback model implemented in the 10xB20 simulations. The softening lengths are $4.2$ ckpc for the DM and gas particles and $1$ ckpc for the BH and star particles, respectively. We dub these simulations 1xB20. In addition, for another sub-sample of four Dianoga cluster halos, simulations were carried out also with a mass resolution 25 times better than in the 1xB20 sample and using the same feedback scheme of B20. In this case, the softening lengths are 1.38 ckpc for DM and gas particles and 0.35 ckpc for star and BH particles. We refer to these simulations as 25xB20. We perform the same lensing analysis outlined above also with these smaller data sets and compare the results to the corresponding subsets of 10xB20 simulations.

Fig.~\ref{fig:examples_sim_feedback_res} shows three convergence maps for the same cluster projection in the 1xB20 and 10xB20, and 25xB20 sets. The maps refer to a source redshift of $z_s=3$. We also show the primary and secondary critical lines in white and yellow, respectively. There are no obvious differences between the three simulations regarding the numbers and sizes of secondary critical lines. This result suggests that the three lenses have similar GGSL probabilities. 

\begin{figure}
   \centering
   \includegraphics[width=1.0\linewidth]{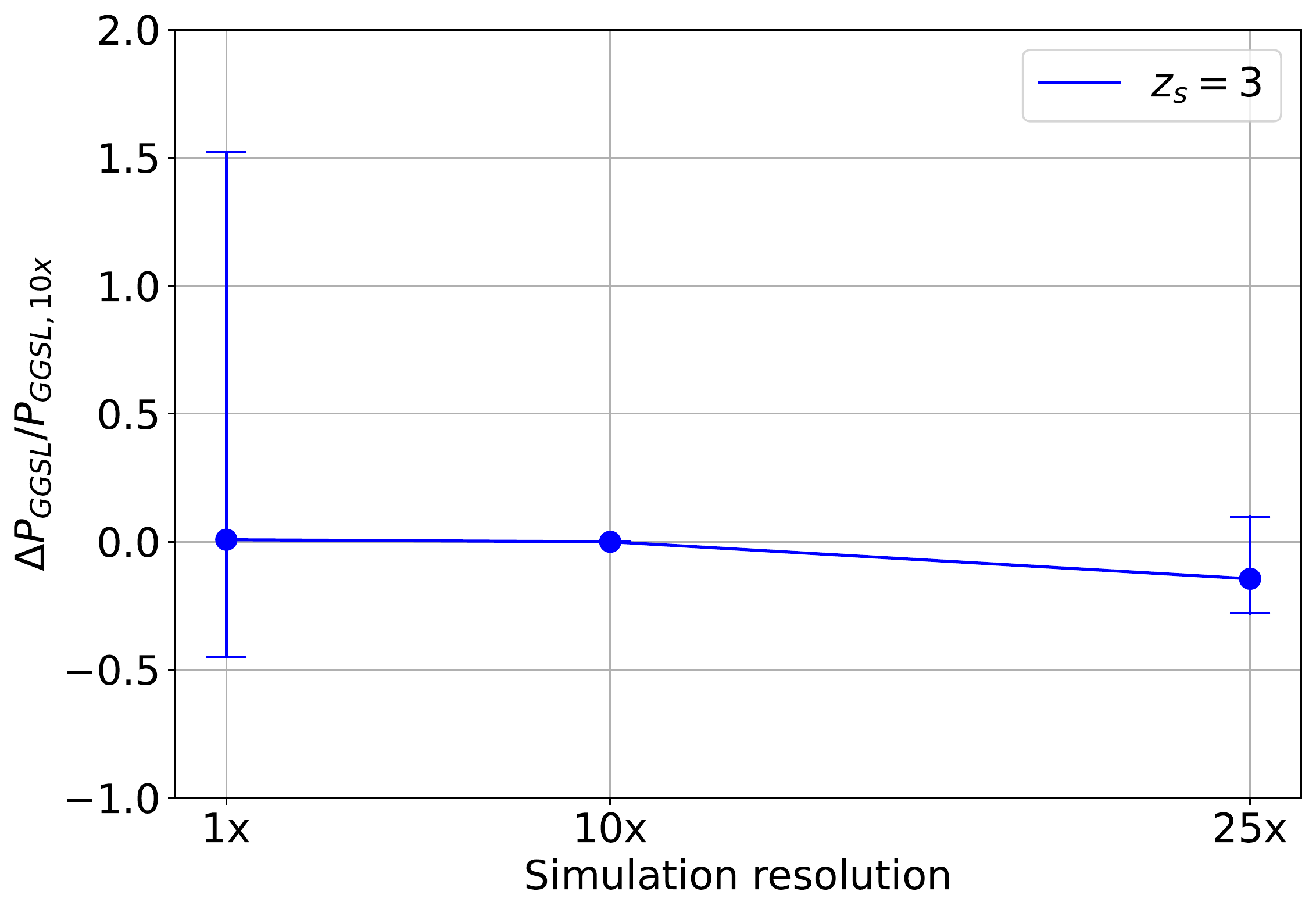}
      \caption{Median relative change of GGSL probability with respect to the 10xB20 set, as a function of the mass resolution in the B20 simulations. The error bars show the 99\% confidence limits.}
\label{fig:mediandiff}
\end{figure}

We find similar results by analyzing the sample of 12 cluster projections at $z\sim 0.4$ obtained from all four clusters available in the 1xB20 and 25xB20 samples. The median relative variation of GGSL probability between these simulation sets and the 10xB20 cluster halos are shown in Fig.~\ref{fig:mediandiff}. Although the error bars are quite large, on average, the GGSL probability is nearly independent on resolution. We show the results for a source redshift of $z_s = 3$. They are similar for other source redshifts. The GGSL probabilities slightly decrease, rather than increase, as a function of mass resolution. 
%Only for $z_s=1$ we find that the GGSL probabilities are smaller by $\sim 25\%$ in the 1xB20 than in the 10xB20 simulation set. However, for this source redshift, the presence (or absence) of only one subhalo contributing to the GGSL cross-section can strongly impact on the results. 
We conclude that the increment of GGSL probability between the 1x (RF18 or R15) and 10xB20 simulations is mostly due to the different feedback schemes implemented in these simulations rather than the higher numerical resolution in the 10xB20 data set.

Our results agree with those of M20, who carried out extensive tests to ensure that the limited resolution of the 1xR15 simulations was not impacting their measurements of the GGSL cross-sections and probabilities. On the contrary, they are inconsistent with those of \cite{2021MNRAS.504L...7R}, based on the {\sc C-EAGLE} simulations, who claims that the GGSL cross-sections measured by M20 are under-estimated by up to a factor of 2. This inconsistency most likely depends on the method employed by \cite{2021MNRAS.504L...7R} to quantify the impact of mass resolution. In our analysis, we consider a set of clusters simulated with three different resolution levels. This is critically important to preserve the appropriate dynamical state of subhalos while changing only the resolution. On the other hand, the low-resolution clusters considered in \cite{2021MNRAS.504L...7R} were generated by re-sampling the particle distributions of higher resolution simulations. 
Under-sampling a high-resolution simulation to a given number of particles is not expected to produce the same results as running a lower-resolution simulation with the same number of particles. 

\subsection{Subhalo contributions to the GGSL cross-sections}
\label{sect:substhetae}
The GGSL probability is an integrated quantity, i.e. it does not inform us on the properties of the individual subhalos, but rather of those that contribute mostly to the GGSL signal. 

We can characterize the subhalos in terms of their equivalent Einstein radii (Eq.~\ref{eq:einstrad}). In practice, not all subhalos have their own critical lines and, in several cases, the critical lines associated to nearby subhalos merge forming larger critical lines enclosing more than one subhalo. Thus, if $N_{\rm crit}$ is the total number of secondary critical lines and $N_{sub}$ is the total number of subhalos in a cluster, we have that $N_{\rm crit}\leq N_{sub}$.

\begin{figure}
   \centering
   \includegraphics[width=8cm]{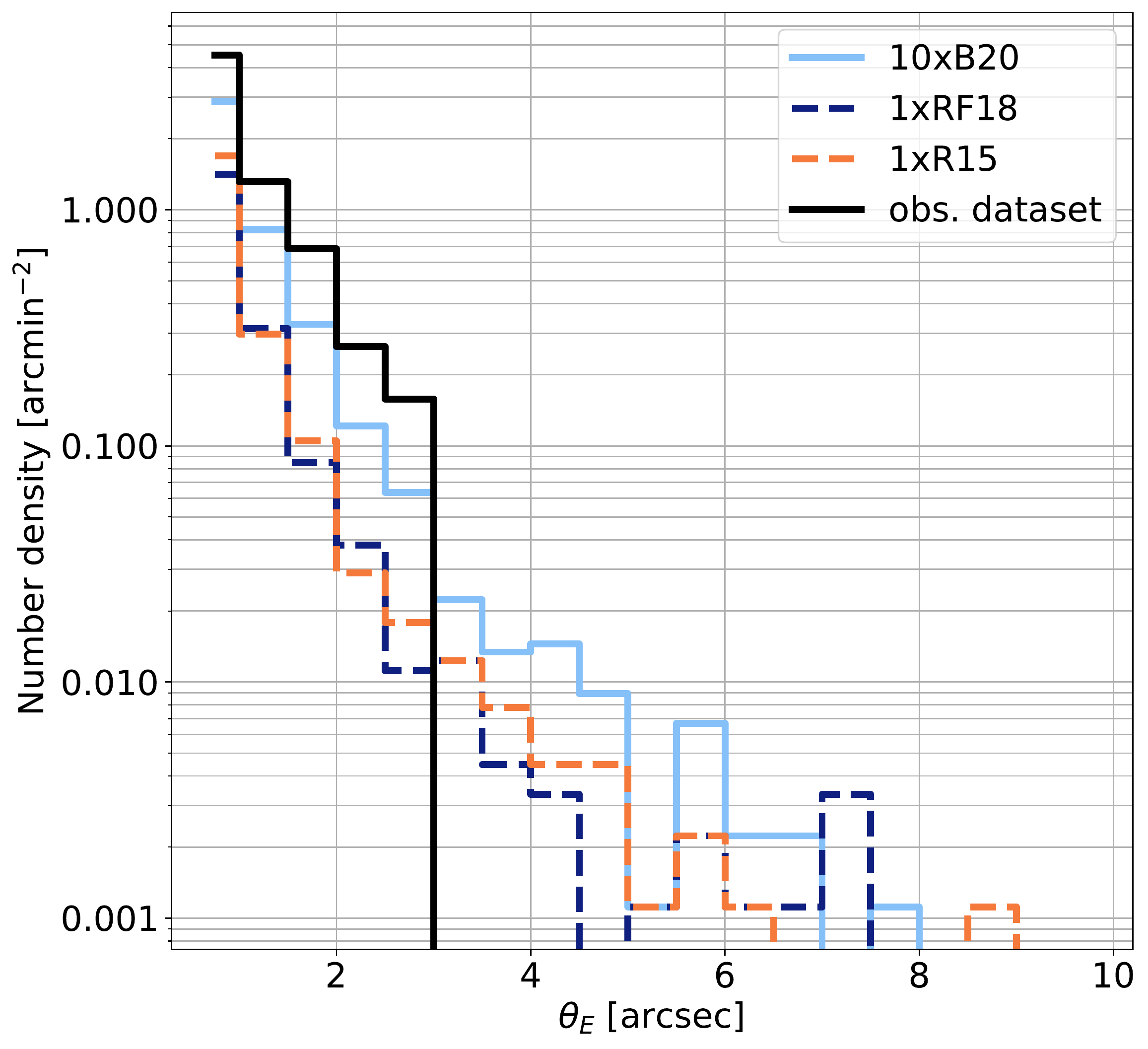}
      \caption{Number density of secondary critical lines as a function of their equivalent Einstein radius $\theta_E$.}
\label{fig:thetaedistr}
\end{figure}

The Einstein radius measures the enclosed projected mass and is sensitive to the subhalo mass profile and to the local background cluster surface density. It is also dependent on the intensity of the cluster shear field. 

In Fig.~\ref{fig:thetaedistr}, we show the average number density of secondary critical lines in the observational and simulation data sets as a function of their Einstein radius. We assume $z_s=6$ in this analysis. For this source redshift, the clusters produce the largest number of secondary critical lines. The results for $z_s=3$ are very similar. In the observational data set (black histogram), the distribution of Einstein radii has a cut-off at $\theta_{E,{\rm cut}}\sim 2.5''$. The number density of critical lines with $\theta_E<\theta_{E,{\rm cut}}$ exceeds that of the 10xB20 sample (light blue histogram) by a factor $\sim 2$. In the case of the 1x simulations (dark blue and orange dashed histograms), the gap with observations is a factor $\sim 7$, consistent with the results of M20.

\begin{figure*}
   \centering
   \includegraphics[width=0.8\linewidth]{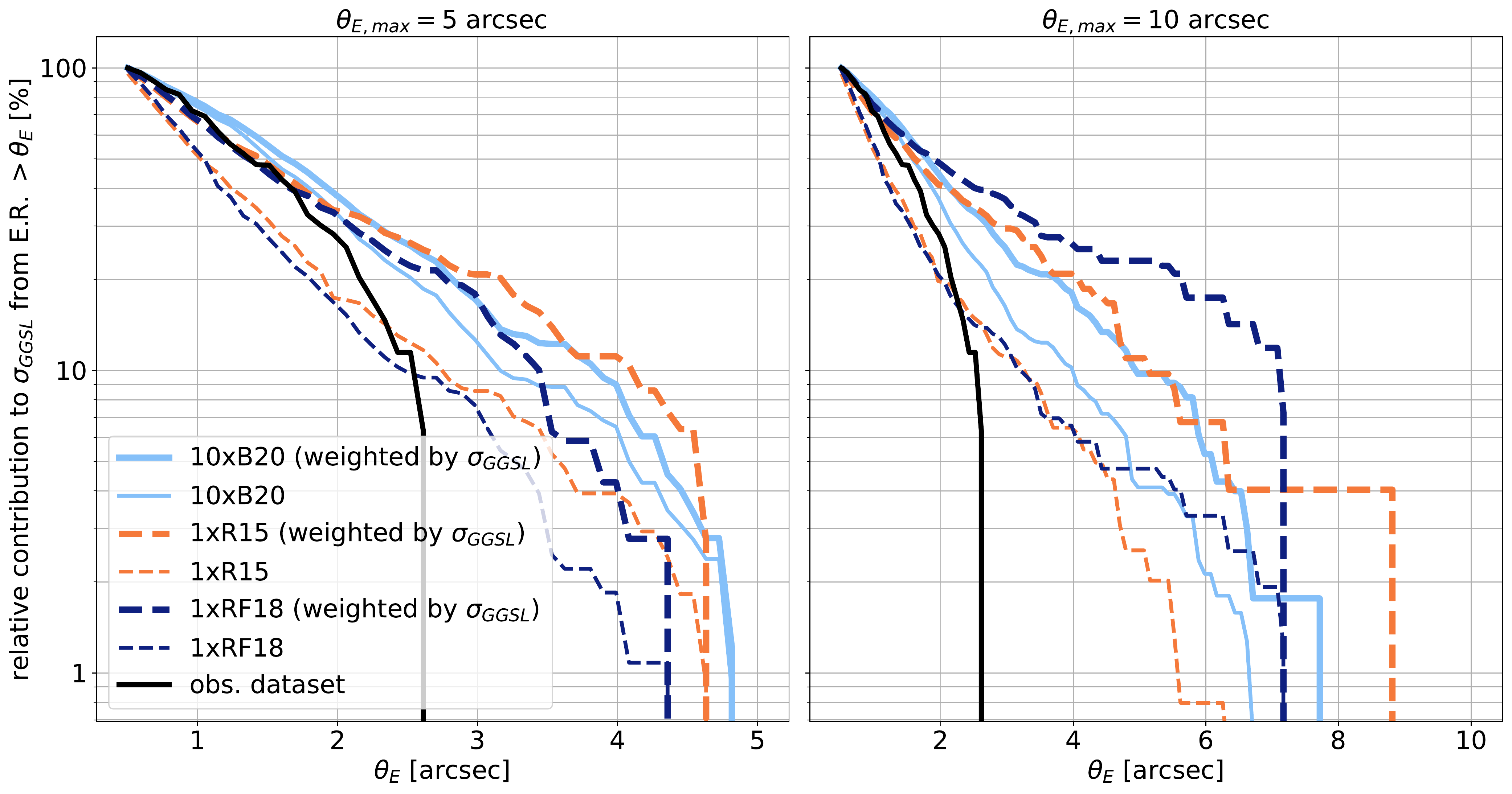}
      \caption{Mean relative contribution to the GGSL cross-section from critical lines with Einstein radii larger than $\theta_E$. The solid black line shows the results for the observational data set. The thin dark blue,  light blue, and orange lines refer to the 10x, 1xRF18, and 1xR15 simulations. The thick lines show the mean over the cluster projections in each data set again, but weighing by their total GGSL cross-sections. The left and right panels show the results for $\theta_{E,{\rm max}}=5''$ and $\theta_{E,{\rm max}}=10''$, respectively.}
\label{fig:thetaecontrib}
\end{figure*}

Interestingly, the Einstein radii distributions of the simulation data sets have tails extending to $\theta_E \sim 8-9''$. As noted earlier, the clusters in the observational data set do not have secondary critical lines with such large extensions as a function of Einstein radius. 

M20 define an upper limit $\theta_{E,{\rm max}}=5''$ for the Einstein radii of secondary critical lines used to compute the GGSL cross-sections \citep[see also ][]{2021MNRAS.504L...7R}. In the left panel of Fig.~\ref{fig:thetaecontrib}, we show the mean relative contribution to the cross-sections from critical lines with Einstein radii larger than a minimum value, averaging over all clusters in each data set. The solid black line shows the results for the observational sample. Nearly $50\%$ of the cross-section is contributed by Einstein radii larger than $\sim 1.3''$. Obviously, this contribution drops to zero for $\theta_E>\theta_{E,{\rm cut}}$. 

There are significant differences between the 1x and 10xB20 data sets. The light blue line for the 10xB20 data set is always above the curve for the observational data set. In particular, critical lines with $\theta_E>\theta_{E,\rm cut}$ give a non-negligible mean contribution to the GGSL cross-section of $\sim 20\%$. The thin dashed dark blue and orange lines for the 1xRF18 and 1xR15 simulations are generally below the curve for the observational data set, indicating that smaller critical lines give a more significant contribution to the GGSL cross-section. However, even in these cases, critical lines with $\theta_E>\theta_{E,\rm cut}$ supply no more than $\sim 10\%$ of the total cross-section.

The thick lines in Fig.~\ref{fig:thetaecontrib} show the same results when we average over the cluster projections in each simulation data set, weighing by their GGSL cross-section. Since each thick curve is above the corresponding thin line, the impact of the most prominent critical lines is more significant in the cluster projections with large GGSL cross-sections. This effect is particularly striking in the 1x simulations but is also substantial for the 10xB20 data set. In fact, for some cluster projections, the critical lines with $\theta_E>\theta_{E,{\rm cut}}$ contribute to more than $50\%$ of the GGSL cross-section.

The right panel of Fig.~\ref{fig:thetaecontrib} shows the same results when the GGSL cross-sections are computed with $\theta_{E,{\rm max}}=10''$. As expected, the contribution of critical lines with $\theta_E>\theta_{E,{\rm cut}}$ is even more substantial in this case. Thus, the choice of $\theta_{E,{\rm max}}$ to define the GGSL cross-section is of great importance. In our calculations, we choose $\theta_{E,{\rm max}}=3''$ to be consistent with the maximum Einstein radius measured in the observational data set. 

\subsection{Large Einstein radii}
\begin{figure}
   \centering
   \includegraphics[width=1.0\linewidth]{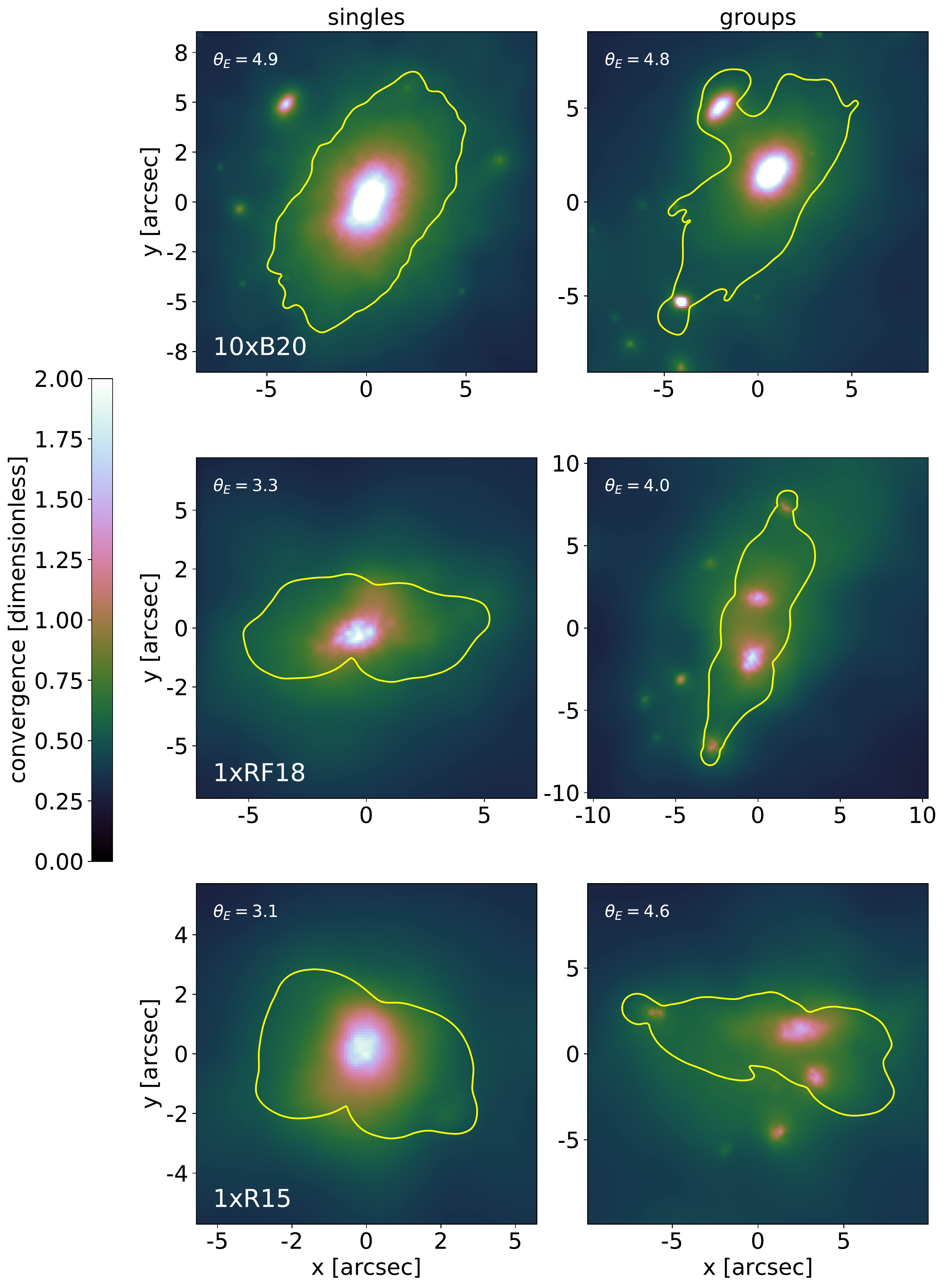}
      \caption{The yellow lines in the left and right panels show examples of critical lines with $\theta_E>\theta_{E,{\rm cut}}$, classified as {\em singles} and {\em groups}, respectively. From the upper to the bottom panels, the critical lines are overlaid onto the convergence maps of simulated clusters in the 10x, 1xRF18, and 1xR15 samples. The value of the corresponding Einstein radius is reported in each panel.}
\label{fig:cl_examples}
\end{figure}

We focus on the critical lines in the simulation sets with $\theta_E>\theta_{E,{\rm cut}}$. As explained earlier, we do not find critical lines with these large sizes in the observational data set and we aim at better understanding their origin. We divide them into two categories. The first includes the so-called {\em singles}, i.e., critical lines that originate from single subhalos. The second comprises the critical lines resulting from mergers of smaller critical lines. They enclose multiple subhalos, and, for this reason, we call them {\em groups}. We show examples of these two categories of critical lines in Fig.~\ref{fig:cl_examples}, for each simulation data set. We overlay the critical lines onto the convergence maps to visualize the mass distribution in the subhalos.

\begin{figure}
   \centering
   \includegraphics[width=1.0\linewidth]{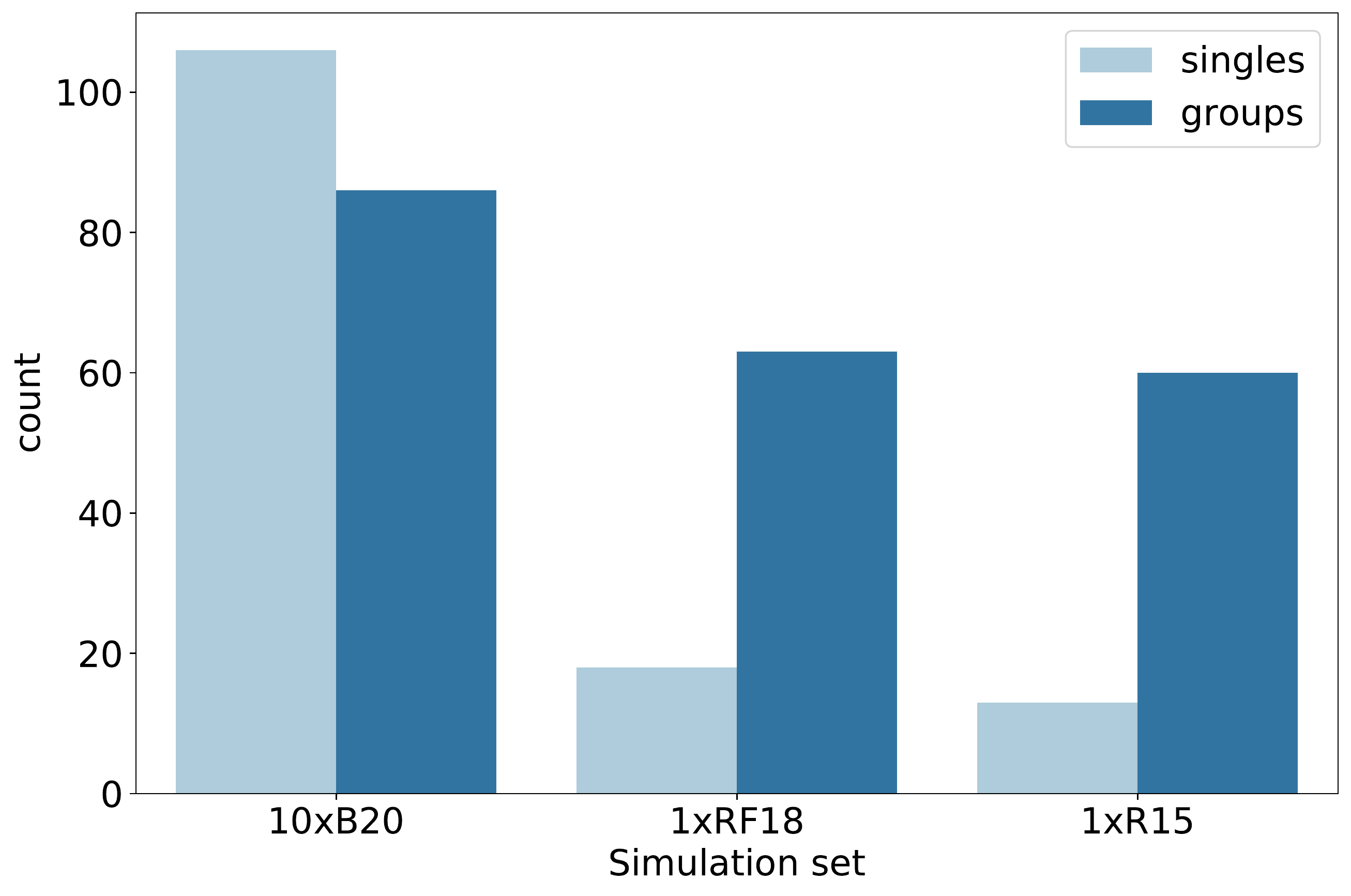}
      \caption{Counts of {\em singles} and {\em groups} in the three simulation data sets. The results refer to all secondary critical lines with $\theta_E>\theta_{E,{\rm cut}}$ for $z_s=1,3$, and $6$.}
\label{fig:cl_classes}
\end{figure}

In Fig.~\ref{fig:cl_classes}, we show the counts of these large critical lines in each of the two categories for all simulation data sets. In the case of the 1xRF18 and 1xR15 simulations, about two-thirds of the critical lines are classified as groups. Only a minority of them are singles. This partition is inverted in the 10xB20 data set, where the singles are more abundant than the groups. 

This difference is interesting because it shows that single galaxy-scale subhalos in the 10xB20 simulations are massive and compact enough to become critical for lensing. Their critical lines have Einstein radii typical of small galaxy clusters and groups. We expect that the dense cluster environments contribute to making cluster galaxies capable of producing strong lensing effects. However, as noted above, Einstein radii as large as $\sim 5''$ are inconsistent with the observations of cluster galaxies.

On the contrary, the few large critical lines in the 1xR15 and 1xRF18 samples are predominantly associated with groups of subhalos, indicating that in most cases, single galaxies would remain sub-critical or develop much smaller Einstein radii. The examples in Fig.~\ref{fig:cl_examples} show that the subhalo mass distribution in the 1x simulations is more diffuse than in the 10xB20 sample. As we pointed out in Sect.~\ref{sect:simulations}, the AGN feedback scheme implemented in these simulations is more efficient than in the 10xB20 simulations. For this reason, gas cooling and star formation occur at lower rates, preventing the formation of stellar cores as dense as those that form in the 10xB20 simulations. In addition, as described by \cite{2012MNRAS.423.3243R}, strong AGN feedback also causes the inner region of the galaxy DM halos to expand and their density profile to flatten.

\subsection{Subhalo compactness}

\begin{figure}
   \centering
   \includegraphics[width=1.0\linewidth]{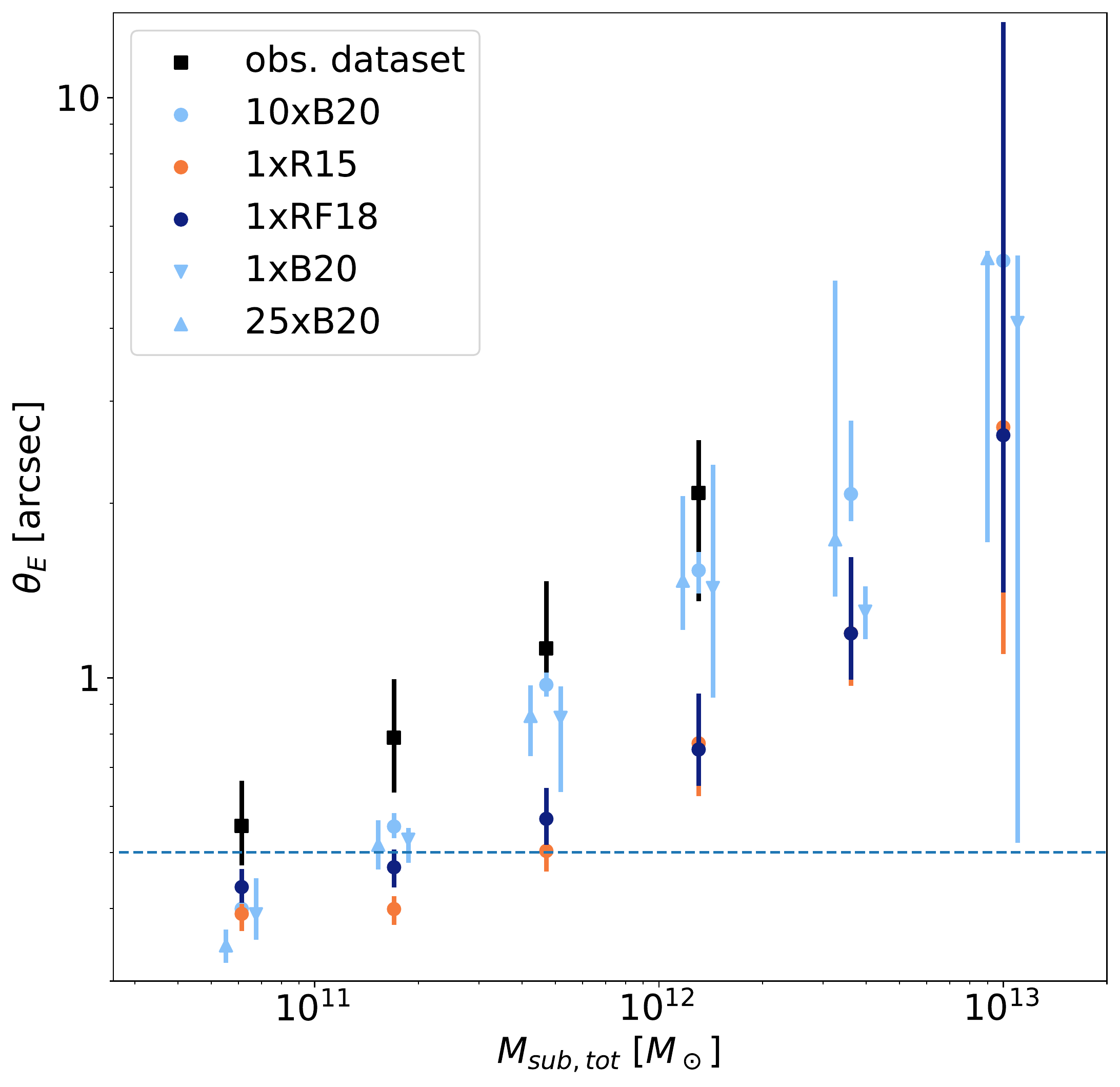}
      \caption{Median Einstein radius as a function of the total subhalo mass. The results refer to a source redshift of $z_s=6$. We use different symbols and colors to display the differences between the data sets. The error bars show the 99\% confidence limits of the median. The horizontal dashed line shows the threshold below which secondary critical lines do not contribute to the GGSL cross-sections in our analysis.}
\label{fig:thetae_msub}
\end{figure}
In Fig.~\ref{fig:thetae_msub}, we show the median Einstein radius of secondary critical lines in bins of increasing subhalo mass for the observational and simulation data sets. For the observational data set, the subhalo masses are given by the {\sc Lenstool} reconstructions, where each subhalo is attached to a cluster galaxy. For the simulation data sets, the particles belonging to each subhalo are identified using the software {\sc Subfind}, as outlined in \cite{ragagnin22}. Since, as discussed earlier, several secondary critical lines contain more than one subhalo, we compute the {\em total} subhalo mass associated with each secondary critical line by adding the masses of all the subhalos it contains.

As expected, the Einstein radius grows as a function of the subhalo mass. Subhalos in the observational data set (black squares and error bars) have Einstein radii systematically larger than their simulated counterparts with similar masses. At masses $\lesssim 2\times 10^{11}\, M_\odot$, the difference amounts to $\sim 50-60\%$, independently of the simulation data set. Interestingly, in this mass limit, the average Einstein radii of simulated subhalos are smaller than the threshold of $0.5''$, below which we do not account for the subhalo contribution to the GGSL cross-section. As the subhalo mass increases, the gap between the B20 simulations and the observations reduces. This result holds independently of the resolution, i.e., the Einstein radii of subhalos in the 1x, 10x, 25x simulations (light blue symbols and error bars) are very similar at fixed subhalo mass. Once more, this result confirms that the mass resolution has a negligible impact on our conclusions. The subhalos in the 1xR15 and 1xRF18 simulations have Einstein radii significantly smaller than those in the observational data at all mass scales.

The observational data set does not contain subhalos with masses $\gtrsim 2 \times 10^{12}\, M_\odot$ producing secondary critical lines. On the contrary, several massive subhalos exist in the simulation data sets. They have large Einstein radii, as discussed in the previous section.

M20 show that, at fixed mass, the maximum circular velocities of subhalos in the 1xR15 simulations are systematically lower than measured in cluster galaxies \citep[e.g.][]{2019A&A...631A.130B}. The maximum circular velocity, $V_{\rm max}$, is often quoted as a proxy for the subhalo compactness. Thus they conclude that simulated subhalos are less compact than their observed counterparts. Consequently, they are less efficient strong lenses and have smaller Einstein radii. \cite{ragagnin22} show that a similar result holds for the subhalos in the 1xRF18 data set. In the 10xB20 (and in the 1xB20 and 25xB20) simulations, the relation between $V_{\rm max}$ and subhalo mass is fully consistent with the 1xRF18 and 1xR15 simulations at masses $M_{sub}\lesssim 10^{11} M_\odot$. On the contrary, at higher masses, the relation is steeper, i.e., the subhalos in the 10xB20 simulations have larger $V_{\rm max}$ than subhalos of equal mass in the 1xRF18 and 1xR15 data sets. Apparently, this behavior brings the 10xB20 simulations in better agreement with the observed $V_{\rm max}-M_{sub}$ relation at these mass scales. However, as shown by \cite{2020A&A...642A..37Bassini}, their AGN feedback model leads to the formation of central galaxies whose stellar masses are too high compared to observations. Thus, the effects of baryons on the inner region of massive galaxies in these simulations is overestimated. They cause the formation of overly massive, compact galaxies that lie on the extrapolation of the observed $V_{\rm max}-M_{sub}$ relation in a mass range rarely populated by real galaxies.  The presence of huge secondary critical lines associated with single subhalos in the 10xB20 data set reflects the same problem. Unrealistically massive subhalos with dense stellar cores are very efficient strong lenses. Thus, their Einstein radii are larger than those of the brightest cluster galaxies.

\cite{2021MNRAS.505.1458B} reports a similar steepening of the $V_{\rm max}-M_{sub}$ relation at high masses in the {\sc Hydrangea/C-Eagle} simulations. The stellar
masses of central galaxies in {\sc C-EAGLE} are 0.3–0.6 dex above their observed counterparts \citep{2017MNRAS.470.4186Bahe,2018MNRAS.479.1125R}. Thus, the {\sc Hydrangea/C-Eagle} likely suffer from the same systematic problem as our 10xB20 simulations. \cite{2021MNRAS.504L...7R} uses these simulations to compute the GGSL probability. They find unsurprisingly, that for the most massive cluster halos in their sample, the integrated GGSL is consistent with observations. If their simulations are similar to ours, it is very likely that a significant contribution to their GGSL cross-section comes from {significantly more massive galaxies that are not present in the observations}, as we report above.  

M20 showed in their Fig. S9 that, by switching off the AGN feedback, the GGSL probability increases by up to one order of magnitude. This effect indicates that we can in fact mitigate the discrepancy between observed and simulated GGSL probability by changing the feedback model. Unfortunately, the price to pay is that the resulting simulated galaxies then have unrealistically high stellar masses and baryon fractions in clear and strong disagreement with observations.

\section{Summary and conclusions}
\label{sect:conclusions}

In this paper, we compare the GGSL probability in numerical hydrodynamical simulations implementing different mass and force resolutions and AGN feedback models. The data sets include seven massive galaxy cluster halos identified in a parent dark-matter-only $\Lambda$CDM cosmological simulation that were re-simulated at higher mass and force resolution, including baryons, starting from the same initial conditions using the zoom-in technique. From their particle distributions at different redshifts, we derived hundreds of projected mass distributions whose lensing properties we study using the ray-tracing technique. We compare the GGSL probability in the simulations with results from the strong lensing mass modeling of four observed galaxy clusters. Three clusters (Abell S1063, MACS J0416.1-2403, and MACS J1206.2-0847) belong to the CLASH and Hubble Frontier Fields samples. They were part of the reference sample of M20. An additional object, PSZ1 G311.65-18.48, included here is part of the {\em Planck} SZ selected cluster sample. We recently modeled these clusters using a novel technique that combines HST imaging and VLT/MUSE spectroscopy data. The method delivers improved mass reconstructions on the scales of cluster galaxies. 

The cluster halos in the 10xB20  data set were simulated with ten-times better mass resolution than in the 1xRF18 and 1xR15 data sets. In addition, they implement an AGN feedback scheme that is less efficient at suppressing gas cooling and star formation. The 1xRF18 and 1xR15 data sets have the same particle masses but different softening lengths.

We summarize our results as follows:
\begin{itemize}
\item independent of the resolution and AGN feedback scheme adopted, the GGSL probabilities in all simulation data sets are lower than in the observed galaxy cluster lenses;
\item the GGSL probability in the higher mass resolution 10xB20 data set is higher than in the 1xRF18 and 1xR15 data sets. The difference in GGSL between these simulations depends on the cluster mass. For cluster halos with masses $M_{200}>10^{15}\,M_\odot$, we measure a GGSL probability higher by a factor $\sim 3$. For lower mass cluster halos, the difference amounts to a factor $\sim 6$;
\item for a sub-sample of the 10xB20 data set, we ran new simulations using the same AGN feedback scheme but lowered the mass resolution by a factor of 10. Compared with the original 10xB20 simulations, we find that the GGSL probability changes only by a few percent. We find similar results also by increasing the mass resolution by a factor $2.5$ compared to the 10xB20 data set. Thus, we exclude that mass resolution strongly impacts the results. On the contrary, the higher GGSL probability in the 10xB20 simulations is due to the less efficient AGN feedback scheme that favors the formation of dense stellar cores and overly massive galaxies;
\item the AGN feedback schemes implemented in the 1xRF18 and 1xR15 simulations have comparable efficiency at suppressing gas cooling and star formation. Despite the larger gravitational softening of the 1xRF18 simulations, the GGSL probabilities in the two data sets are very similar. Thus, force resolution also has a small impact on the GGSL results;
\item we quantify the size of the critical lines used to compute the GGSL cross-sections and probability using their equivalent Einstein radius, $\theta_E$. The distribution of Einstein radii in observed galaxy clusters is truncated at $\theta_{E,{\rm cut}}\sim 2.5''$. On the contrary, simulated cluster subhalos develop critical lines with Einstein radii as large as $\sim 8''$. In the 10xB20 simulations, the critical lines with $\theta_E>\theta_{E,{\rm cut}}$ on average contribute at least $20\%$ of the GGSL cross-section. In some cases, particularly in the cluster lenses with the largest GGSL cross-sections, their contribution is much more substantial ($\gtrsim 50\%$);
\item most of the critical lines with $\theta_E>\theta_{E,{\rm cut}}$ in the 1xRF18 and 1xR15 enclose multiple subhalos. Thus, they are the result of the mergers of smaller critical lines. Single subhalos in these simulations are unable to produce such large critical lines. On the contrary, in the 10xB20 simulations, more than $50\%$ of the critical lines with $\theta_E>\theta_{E,{\rm cut}}$ enclose a single subhalo. Thanks to their less efficient AGN feedback scheme, these simulations form compact and massive subhalos to become super-critical for strong lensing. Such extended critical lines are inconsistent with observations and are associated with overly massive subhalos.
\end{itemize} 
Based on these results, we re-affirm the tension, previously reported in M20, between observations of  GGSL in galaxy clusters and theoretical expectations in the framework of the $\Lambda$CDM cosmological model. Observed cluster galaxies are stronger lenses than subhalos in numerical hydrodynamical simulations. It is unclear if we can fully resolve this discrepancy by changing the nature of dark matter or improving the baryonic physics treatment in the simulations. An AGN feedback scheme that implies more efficient gas cooling and star formation can reduce the gap with observations in terms of GGSL probability. However, the demographics of the subhalos producing GGSL events are significantly different. A small number of overly massive subhalos in these simulations provides a significant fraction of the GGSL cross-section. On the contrary, lower mass subhalos, the more significant contributors to the GGSL cross-sections in observed galaxy clusters, are inefficient strong lenses in the simulations because, at fixed mass, they are less compact than their observed counterparts. 

We conclude that current numerical simulations in the $\Lambda$CDM cosmological model have difficulty in  reproducing the stellar mass function and the galaxy internal structure simultaneously. We emphasize that GGSL in galaxy clusters can be a promising method to investigate the consistency of galaxy and star formation models with observations. We look forward to testing numerical simulations in dark matter models different from CDM and implementing alternative AGN feedback schemes.

\begin{acknowledgements}
Simulations have been carried out using MARCONI at CINECA (Italy), with CPU time assigned through grants ISCRA B, and through INAF-CINECA and University of Trieste – CINECA agreements; at the Tianhe-2 platform of the Guangzhou Supercomputer Center by the support from the National Key Program for Science and Technology Research and Devel- opment (2017YFB0203300); using MENDIETA Cluster from CCAD-UNC, which is part of SNCAD-MinCyT (Argentina). 
We acknowledge financial contributions by PRIN-MIUR 2017WSCC32 "Zooming into dark matter and proto-galaxies with massive lensing clusters" (P.I.: P.Rosati), INAF ``main-stream'' 1.05.01.86.20: "Deep and wide view of galaxy clusters (P.I.: M. Nonino)" and INAF ``main-stream'' 1.05.01.86.31 "The deepest view of high-redshift galaxies and globular cluster precursors in the early Universe" (P.I.: E. Vanzella).
Part of this work was supported by the German
\emph{Deut\-sche For\-schungs\-ge\-mein\-schaft, DFG\/} project number Ts~17/2--1.
AR acknowledges support by  MIUR-DAAD contract number 34843  ``The Universe in a Box''. SB acknowldges partial financial support from the INDARK INFN grant. MV is supported by the Alexander von Humboldt Stiftung and the Carl Friedrich von Siemens Stiftung. MV and KD acknowledge support by the Deutsche Forschungsgemeinschaft (DFG, German  Research  Foundation)  under  Germany's  Excellence Strategy - EXC-2094 - 390783311. KD also acknowledges support through the COMPLEX project from the European Research Council (ERC) under the European Union’s Horizon 2020 research and innovation program grant agreement ERC-2019-AdG 882679. We are especially grateful for the support by M. Petkova through the Computational centre for Particle and Astrophysics (C$^2$PAP). 
PN acknowledges the Black Hole Initiative (BHI) at Harvard University, which is supported by grants from the Gordon and Betty Moore Foundation and the John Templeton Foundation, for hosting her. GBC acknowledges the Max Planck Society for financial support through
the Max Planck Research Group for S. H. Suyu and the academic support
from the German Centre for Cosmological Lensing.
\end{acknowledgements}

% WARNING
%-------------------------------------------------------------------
% Please note that we have included the references to the file aa.dem in
% order to compile it, but we ask you to:
%
% - use BibTeX with the regular commands:
%   \bibliographystyle{aa} % style aa.bst
%   \bibliography{Yourfile} % your references Yourfile.bib
%
% - join the .bib files when you upload your source files
%-------------------------------------------------------------------
\bibliographystyle{aa} 
\bibliography{bibliography.bib} 

\end{document}